\documentclass[12pt]{article}
\usepackage{amsmath}
\usepackage{amsfonts}
\usepackage{mathtools}
\usepackage{bigints}
\usepackage{color}
\usepackage{tabu}
\usepackage[final]{graphicx}
\usepackage{subcaption}
\usepackage[letterpaper, portrait, margin=1in]{geometry}
\usepackage{setspace}
\usepackage{authblk}
\usepackage{float}
\restylefloat{table}
\usepackage[round]{natbib}
\usepackage{enumitem}
\usepackage{multirow}
\usepackage{listings}
\usepackage{lipsum}
\usepackage{chngcntr}

\newcommand{\beq}{\begin{equation}}
\newcommand{\eeq}{\end{equation}}
\newcommand{\beqn}{\begin{eqnarray}}
\newcommand{\eeqn}{\end{eqnarray}}

\usepackage{flexisym}
\usepackage{amsthm}
\theoremstyle{plain}

\usepackage{amsfonts}

\usepackage[linesnumbered,ruled,vlined]{algorithm2e}
\DeclareMathOperator*{\argmax}{arg\,max}

\usepackage{bm}
\usepackage{hyperref}

\begin{document}
\title{Bayesian Spatial Analysis of Hardwood Tree Counts in Forests via MCMC\\}
\author[1]{Reihaneh Entezari\thanks{entezari@utstat.utoronto.ca}}
\author[1,2]{Patrick E. Brown\thanks{patrick.brown@utoronto.ca ~~  web:~http://pbrown.ca}}
\author[1]{Jeffrey S. Rosenthal\thanks{jeff@math.toronto.edu ~~  web:~http://probability.ca/jeff/}}
\affil[1]{Department of Statistical Sciences, University of Toronto}
\affil[2]{Centre for Global Health Research, St Micheal's Hospital}
\maketitle

\begin{abstract}
In this paper, we perform Bayesian Inference to analyze spatial tree count data from the Timiskaming and Abitibi River forests in Ontario, Canada. We consider a Bayesian Generalized Linear Geostatistical Model and implement a Markov Chain Monte Carlo algorithm to sample from its posterior distribution. How spatial predictions for new sites in the forests change as the amount of training data is reduced is studied and compared with a Logistic Regression model without a spatial effect. Finally, we discuss a stratified sampling approach for selecting subsets of data that allows for potential better predictions.
\end{abstract}

\section{Introduction}

\subsection{The forest inventory problem}


The forest industry has a significant impact on the economy of countries such as Canada, making forests an important financial asset. The monetary value of forest assets are mainly determined by their timber, the value of which depends on different features of trees such as size, species, age, defects, etc. Tree species have different types of wood with different qualities, and hence influence the timber value. 

Tree species have two main categories, hardwood (deciduous) trees and softwood (coniferous) trees, with hardwood trees generally having wider leaves that are lost annually, while softwood trees have smaller leaves and retain their leaves throughout the year. Hardwood trees provide much longer lasting wood compared to softwood trees, with slower growth rates which makes them more expensive compared to softwood. Hence, knowing the number of hardwood trees in a forest is valuable information. Collecting data on forests requires hiring workers to travel to different sites around the forests and measure the quantities needed, which can be costly and time consuming.

Remote sensing technologies can overcome this issue. Although they are cheap and efficient and can cover a wide range of geographical areas, they can suffer from lack of accuracy. Geostatistical models are powerful tools for analyzing and predicting such spatial data, and can be used to calibrate remotely sensed data \citep[see][]{calibration}. Existing literatures by \cite{environ1,environ2,environ3} are examples of the importance of statistical models for spatial analysis. The focus of this paper will also be to take advantage of statistical tools to predict the number of hardwood trees using geostatistical models that take into account the spatial factor.





\subsection{Model-based geostatistics}

In the past few decades, spatial statistics has become an established field of statistics with well developed models applied to many real-world problems. Conventional geostatistical models for Gaussian spatial data were first popularized by \cite{matheron} and later on built upon by \cite{cressie}. The generalization of these models for non-Gaussian data were introduced by \cite{diggle}.  

Let $Y_i$ be the observed spatial data at location $s_i$, with arbitrary distribution $f$ that has mean $\lambda$ and possible additional parameters $\gamma$. Consider $X(s_i)$ as the covariates at location $s_i$. Modelling this data with the Generalized Linear Geostatistical Model (GLGM) described in \cite{diggle} and \cite{diggle2007}, will be as following:

\begin{align}
\begin{split}
Y_i | U(s_i) & \sim f[\lambda(s_i),\gamma]\\
g[\lambda(s_i)] & = \mu + \beta X(s_i) + U(s_i)
\end{split}
\label{geomodel}
\end{align}
where $g(.)$ is the link function (i.e. logit or log). Here $U(s)$ is a Gaussian random field $U$ evaluated at location $s$, which is characterized by the joint multivariate normal distribution:
\begin{gather*}
[U(s_1),...,U(s_N)]^T \sim MVN(0,\Sigma)
\end{gather*}
where the elements of $\Sigma$ are defined by a spatial correlation function $\rho$ as
\begin{gather*}
\Sigma_{ij} = \text{cov}[U(s_i),U(s_j)] = \sigma^2 \rho(||s_i-s_j||/\phi, \nu)
\end{gather*}
where $\phi$ is a range parameter and $\nu$ is a vector of other possible parameters. The range parameter $\phi$ controls the rate at which the correlation decreases with distance. There are many possible parametric functions for $\rho$, with Mat\'ern correlation function \citep[see][]{matern} being the most commonly used. The Mat\'ern correlation is defined as:
\begin{gather}
\rho(h;\phi,\kappa) = \frac{1}{2^{\kappa-1} \Gamma(\kappa)}  \Bigg(\frac{||h||}{\phi}\Bigg)^\kappa K_\kappa\Bigg (\frac{||h||}{\phi}\Bigg),
\label{matern}
\end{gather}
where $\Gamma(.)$ is the gamma function and $K_\kappa(.)$ is the modified Bessel function of the second kind of order $\kappa >0$ ($\kappa$ being a shape parameter). This function is particularly interesting, as it is flexible in the differentiability of the Gaussian process $U(s)$ by adjusting $\kappa$ \citep{matern}. 

In the case where the data is Gaussian, Maximum Likelihood Estimates (MLEs) can be used as point estimates for the model parameters. However, when the data is non-Gaussian, because of the unobserved latent variables $U$ present in the model, the likelihood function becomes intractable and it is difficult to calculate the MLEs. Performing Bayesian inference on these models via Markov Chain Monte Carlo (MCMC) methods \citep{gelman,radu_rosenthal} has many advantages (as discussed in \cite{diggle}). The Integrated Nested Laplace Approximation (INLA) algorithm introduced by \cite{inla}, is an alternative to MCMC for Bayesian Inference on latent Gaussian models. However this method has some drawbacks as it approximates marginal posterior distributions rather than joint posterior distributions. There are facilities in the R-INLA software for producing approximate joint posterior samples, but the properties of these samples have yet to be explored.

In this paper, we will analyze the spatial hardwood tree count data collected from the Timiskaming \& Abitibi River forests in Ontario, Canada. Our analysis is constructed in a Bayesian framework for a binomial geostatistical model to predict the proportion of hardwood trees from remotely sensed elevation and vegetation data. For posterior simulations, we implement an MCMC method using the Langevin-Hastings \citep[see][]{langevin} and the Random-Walk Metropolis Hastings \citep[see][]{randomwalk, optimal} algorithms. By reducing the amount of training data fitted to the model, and predicting for the same validation set, we are able to answer questions related to the accuracy of predictions given small amounts of ground truth data collected. We will show that with training data size as small as 10 spatial locations, despite the increase in uncertainty, the true number of hardwood trees lies within a 95\% prediction interval. This conclusion is very valuable as it will significantly reduce costs of collecting ground truth data. We will also compare our results with the logistic regression model where there is no spatial effect. Furthermore, we explore a stratified sampling approach in choosing the training data that will show a potential improvement in predictions.

The paper is organized as follows. The spatial data from the Timiskaming \& Abitibi River Forests are described in section \ref{data}. Section \ref{methods} describes the geostatistical model used for our data and the MCMC algorithm applied to perform Bayesian Inference. In addition, we explain our stratified approach and describe the measurements we will use to compare and assess predictions. Section \ref{results} discusses the numerical results from fitting the data, where comparisons are also made with the Logistic Regression. At last, we summarize our results in Section \ref{discussion}. The Appendix includes results from different simulations.

%

\begin{figure}[h]
\centering
\begin{subfigure}{.45\textwidth}
  \centering
  \includegraphics[width=1\linewidth]{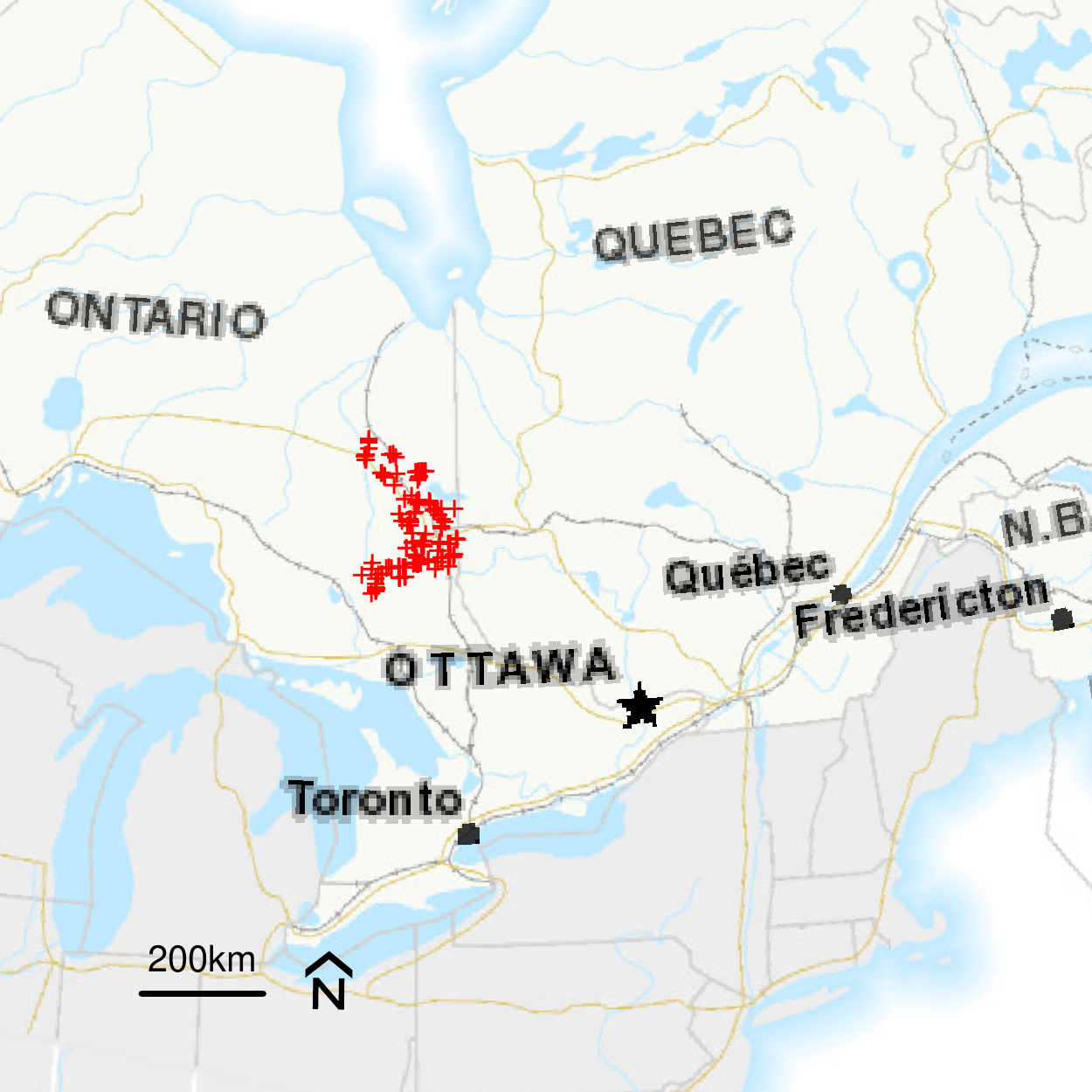}
  \caption{Sample locations}
  \label{fig:forests}
\end{subfigure}
\begin{subfigure}{.45\textwidth}
  \centering
 \includegraphics[width=1\linewidth]{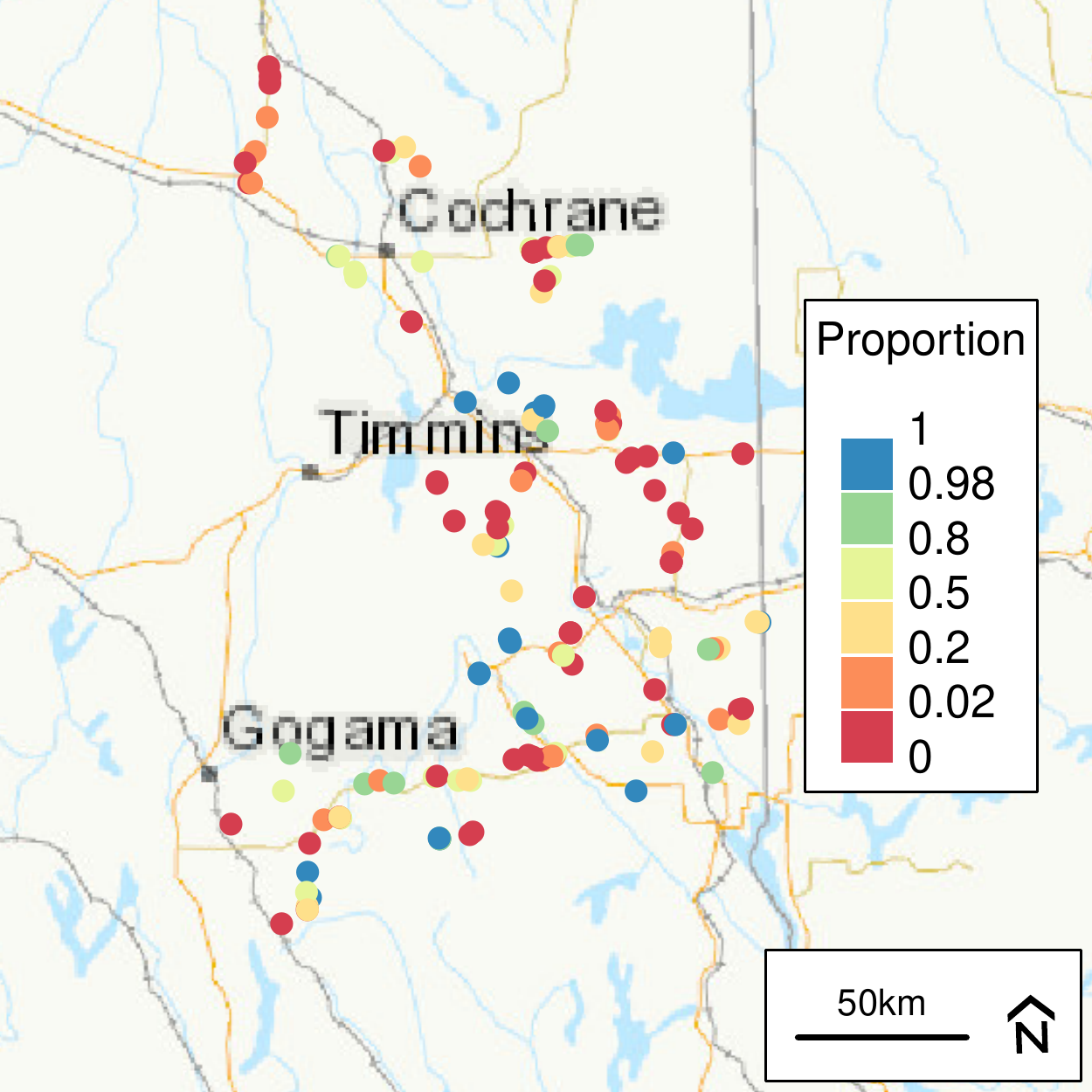}
  \caption{Proportions of hardwood trees}
  \label{fig:hardwoods}
\end{subfigure}
\caption{Locations of 162 forest plots in the Timiskaming and Abitibi River Forests.}
\label{fig:forests_pic}
\end{figure}

\subsection{Description of Data}\label{data}
The Timiskaming and Abitibi River forests are geographically located next to one-another in northern Ontario, Canada. The First Resource Management Group Inc. has provided detailed data from 162 individual forest plots inside these adjacent forests. Each forest plot is $11.28\text{m}$ in radius to provide a $400\text{m}^2$ surface. The geographical locations of these 162 sites are shown in Figure \ref{fig:forests_pic}. 



The data from each site consists of information on the total number of trees, whether each tree is living or dead, and the species of each tree. Figure \ref{fig:hardwoods} shows the proportion of live trees which are hardwood from the 162 sites. As can be seen, many sites have no hardwood trees and such sites are scattered throughout the forests.

\begin{figure}[h]
\centering
\begin{subfigure}{.45\textwidth}
  \centering
  \includegraphics[width=0.9\linewidth]{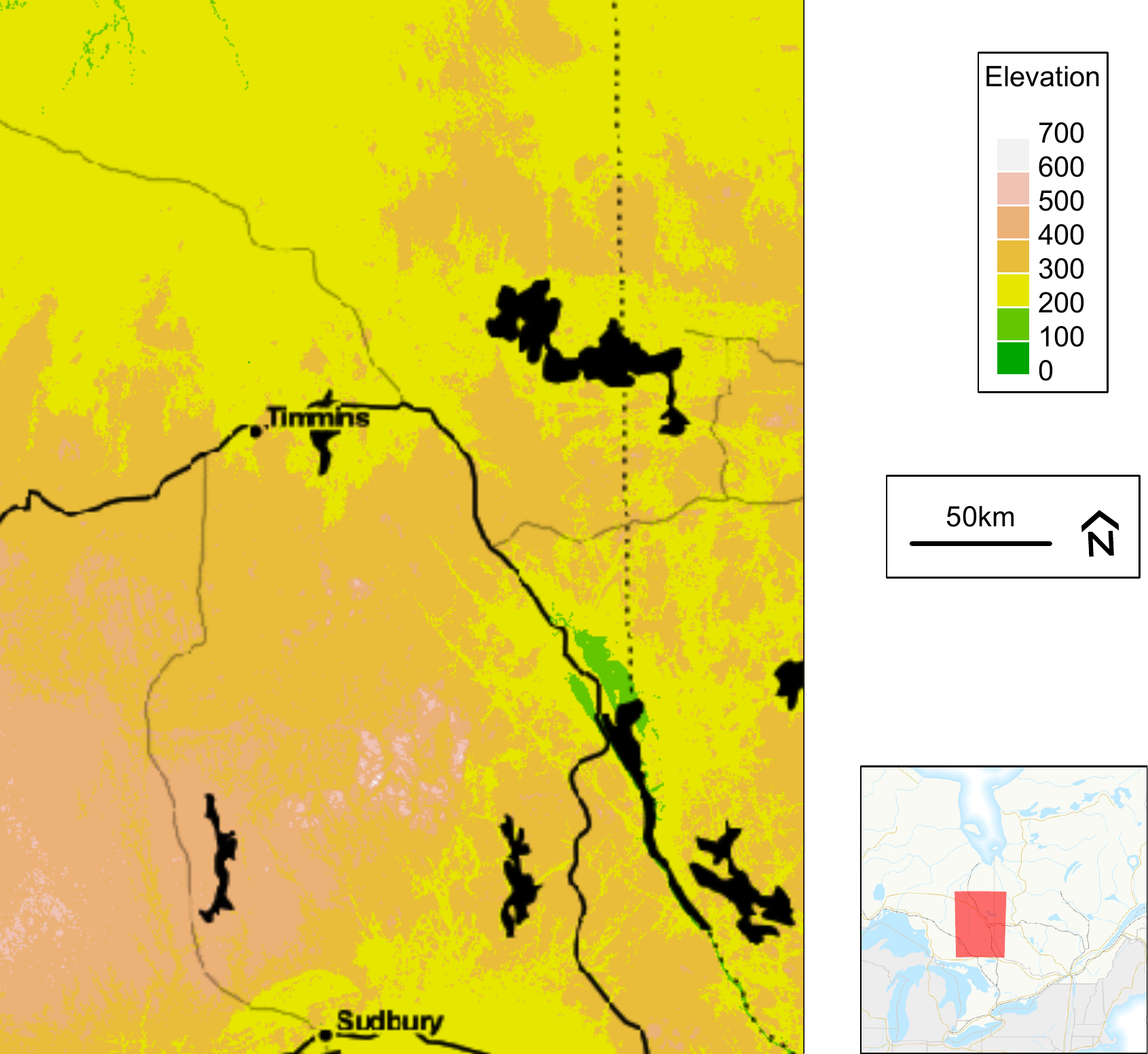}
  \caption{Elevation}
  \label{fig:elev}
\end{subfigure}
\begin{subfigure}{.45\textwidth}
  \centering
  \includegraphics[width=0.9\linewidth]{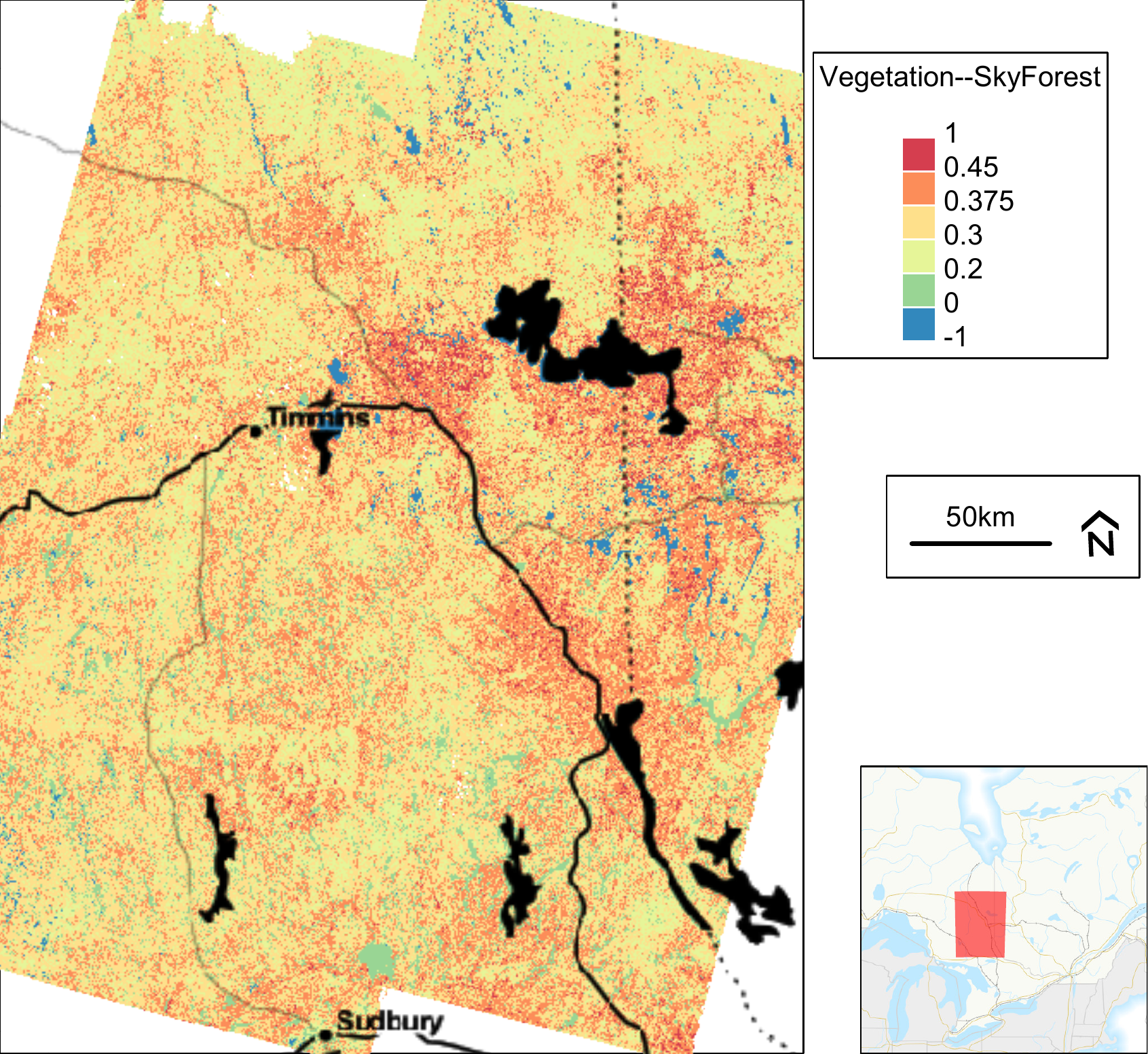}
  \caption{SkyForest$^{\text{TM}}$ vegetation index}
  \label{fig:timisk}
\end{subfigure}
\caption{Elevation \& Vegetation index around the Timiskaming and Abitibi River Forests (Background \copyright \href{http://stamen.com}{Stamen Design}).}
\label{fig:elev_timisk}
\end{figure}


The remotely sensed data considered includes elevation values from satellite data provided by the SRTM program (Figure \ref{fig:elev}). A measure of forest vegetation was provided by the First Resource Management Group Inc. using the proprietary remote sensing technology ``SkyForest$^{\text{TM}}$", which is shown in Figure \ref{fig:timisk}. This vegetation measure is predicted by SkyForest$^{\text{TM}}$ across the forest landscape by selecting an arithmetic transformation of spectral bands (ATSB) from a candidate list of ATSBs. The ATSBs are constructed similarly to well known vegetation indices such as the Normalized Difference Vegetation Index (NDVI), with some of them being multi-temporal. It is thus expected that hardwood trees are located where this measure is high.\\

In the next section, we will describe the geostatistical model for our dataset, along with the steps taken to perform a Bayesian analysis. \\

\section{Methods}\label{methods}
\subsection{Logistic Regression}
Before describing the full geostatistical model for our data, a simple Logistic Regression model with binomial response will be outlined. Consider $Y_i$ to be the count of hardwood trees in forest plot $i$, and write $Y_i \sim \text{Binom}(n_i,p_i)$, where $n_i$ is the total number of live trees at site $i$ ($s_i$) and $p_i$ is the probability of a tree in plot $i$ being hardwood. 
Elevation and the SkyForest$^{\text{TM}}$ index are covariates in the model. The SkyForest$^{\text{TM}}$ covariate is treated as a linear effect with change point at 0.3 (approximately its average value), giving some additional flexibility to this covariate in the regression model. The elevation values are also centered at the average value of about 320. For computational reasons, we normalize the covariates by dividing by the standard deviation. The model is:
\begin{gather}
\begin{split}
Y_i  \sim \text{Binom}(n_i,p_i) ~~~~i=1,...,162\\
\log \Bigg( \frac{p_i}{1-p_i} \Bigg ) =  X(s_i) \beta ~~~~~ 
\end{split}
\label{glm}
\end{gather}
Writing $A(s)$ as the SRTM-measured altitude at location $s$ and $V(s)$ as the SkyForest$^{\text{TM}}$ vegetation index, the normalized vector of covariates $X(s)$ is constructed by:
\begin{align*}
X_1(s) &= 1 \\
X_{2}(s) & = \frac{A(s) - 320}{50} \\
X_{3}(s) &= \frac{\text{min}(V(s)-0.3,0)}{0.05}\\ 
X_{4}(s) &= \frac{\text{max}(V(s)-0.3,0)}{0.05}
\end{align*}

\subsection{The geostatistical model}

Spatial dependence in the prevalence of hardwood trees should be expected as sites in the forests close to one another may benefit from the same soil, weather, etc, and hence may have similar tree types. Thus we expect a geographical effect to play an important role in explaining such data with a more sophisticated model such as the Generalized Linear Geostatistical Model (GLGM). 
A geostatistical model for our spatial data will have an extra spatial term $U(s)$ and an independent term $Z$ compared to the model in (\refeq{glm}), resulting:
\begin{gather}
\begin{split}
Y_i \sim \text{Binom}(n_i,p_i) ~~~~i=1,...,162\\
\log \Bigg( \frac{p_i}{1-p_i} \Bigg ) = t_i =  X(s_i) \beta   + U(s_i) + Z_i\\
\label{spatial}
\end{split}
\end{gather}
where $$Z_i ~{\stackrel{i.i.d.}{\sim}}~ N(0,\tau^2),$$ $$U(s) \sim N(0,\sigma^2),$$ $$\text{cov}(U(s+h),U(s))=\sigma^2\rho(||h||;\phi,\kappa)$$


This model is equivalent to (\ref{geomodel}) where $f$ is Binomial and $g$ is a logit link function.

\subsection{Random Sampling vs Stratified Sampling}
\begin{figure}[h]
\centering
\begin{subfigure}{.45\textwidth}
  \centering
  \includegraphics[width=1\linewidth]{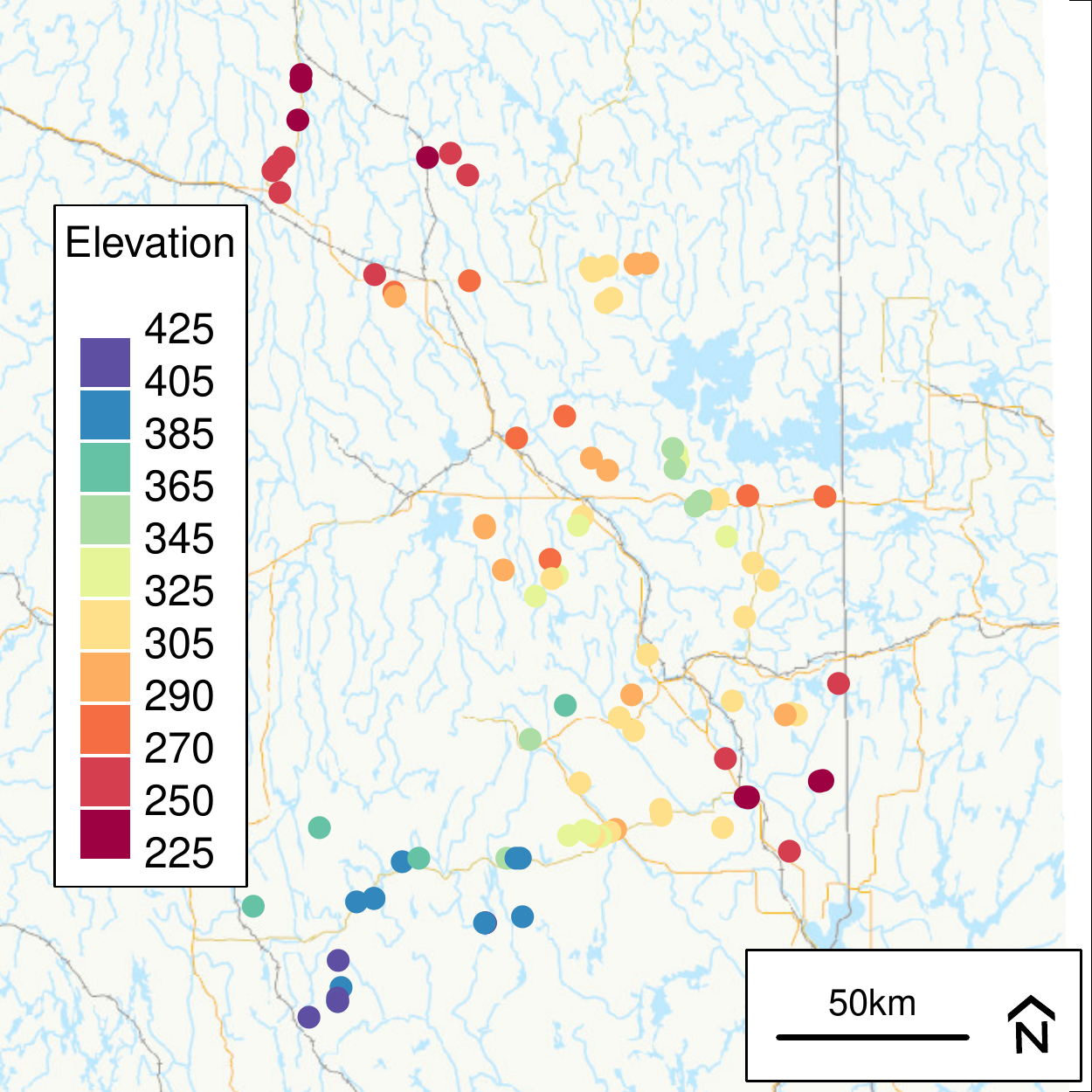}
  \caption{Elevation for 100 training data}
  \label{elev100}
\end{subfigure}
\begin{subfigure}{.45\textwidth}
  \centering
  \includegraphics[width=1\linewidth]{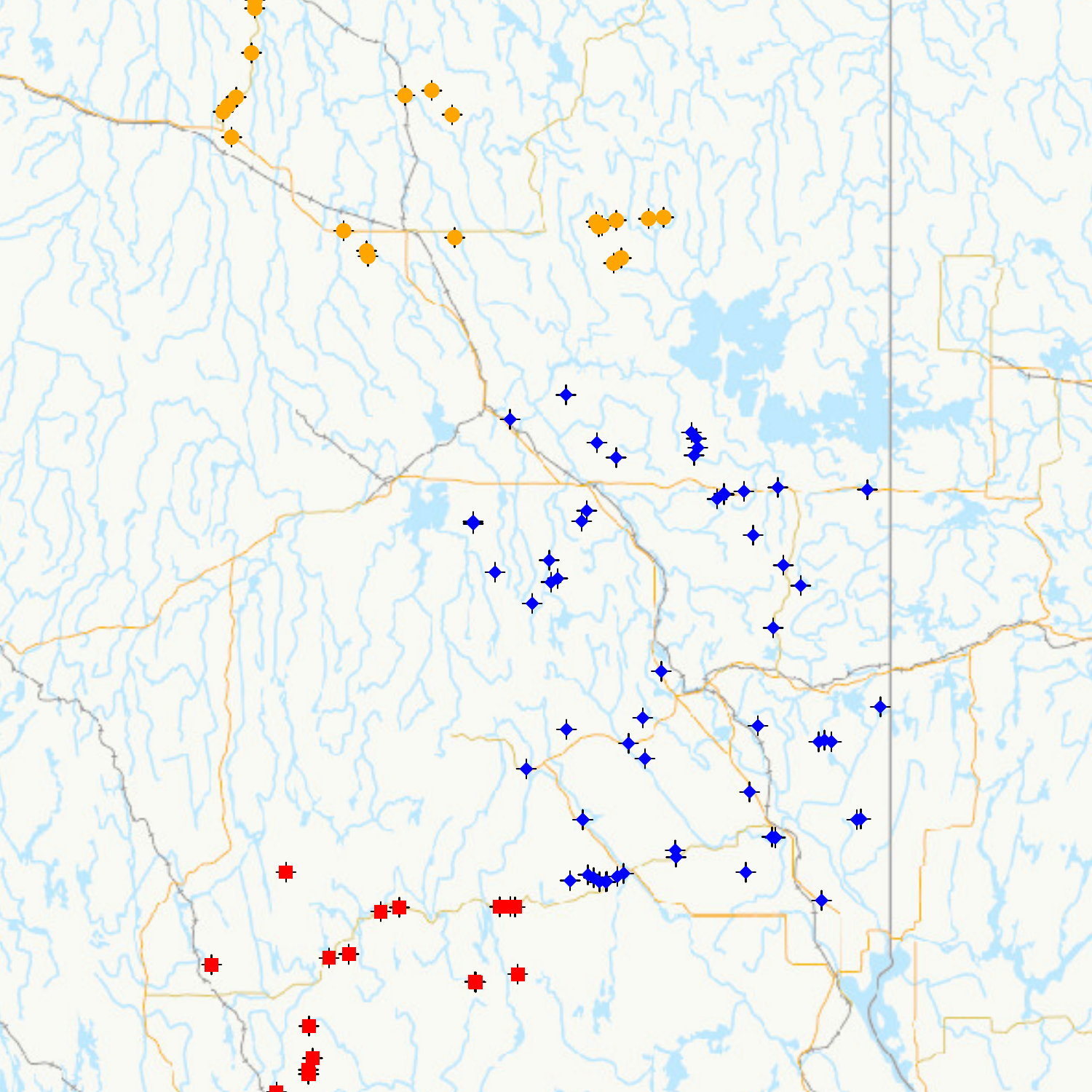}
  \caption{Stratified regions}
  \label{strat100}
\end{subfigure}
\caption{Plots of elevation from 100 training data, along with the plot of stratified regions.}
\label{elev-strat}
\end{figure}

For our analysis, we explore reducing the size of the training data fitted to the model, to observe and examine the trade-off between prediction accuracy and costs of collecting ground truth data. More specifically, if we were only given data from 25 or 10 plots on the ground, could useful predictions still be made? To answer this question, the 162 plots in the dataset were divided into 100 training and 62 validations sets. Keeping the 62 validation set fixed, we examine the performance of results generated by fitting 100, 25, and 10 training data to the model. For this purpose, we can do this by two different approaches, 1) choosing random subsets of data and 2) choosing stratified subsets of data. Since the spatial data is correlated, choosing the subset of data with a stratified approach should be expected to improve the results, as it can force the training plots to be as scattered as possible. Both elevation and vegetation covariates are taken into account for choosing the 25 and 10 dataset from the 100. Hence, we begin by looking at the elevation from all the 100 training data (first simulation) as shown in Figure \ref{elev100}.

The 100 plots are stratified into three groups as shown in Figure \ref{strat100}, and both location of the points as well as elevation values are taken into account equally. Keeping the proportion of the data from each strata constant, we systematically sample 25 plots from the 100 by sorting the vegetation index in each strata and taking every $j-th$ element depending on the number of data needed (similarly for the 10 data points from the 25). The Results section will explore how stratified sampling can (potentially) improve prediction accuracy with smaller training data fitted to the model, compared to random sampling.

\subsection{Inference}\label{bespoke_mcmc}
We will apply a Bayesian approach to the model in (\refeq{spatial}), and this methodology will be referred to as the Bayesian Generalized Linear Geostatistical Model (BGLGM).
Let $\beta^T = (\beta_0,\beta_1,\beta_2,\beta_3)$, $\theta^T =(\sigma^2, \phi,\tau)$, and $t^T = (t_1,...,t_n)$ with $t_i=X(s_i) \beta + U(s_i) + Z_i$, be the three sets of parameters. We treat $\kappa$ as fixed at $1.5$, since it is not of direct interest and according to \cite{zhang}, not all the parameters ($\sigma^2$, $\phi$, and $\kappa$) are consistently estimable. We define priors for each parameter as
\begin{gather*}
\theta \sim \pi_1(.) ~~~~\& ~~~~ \beta|\theta \sim \pi_2(.) = N(\mu,\sigma^2\Omega) ~~~~\& ~~~~ t|\beta,\theta \sim \pi_3(.) = MVN(X\beta,\Sigma(\theta))
\end{gather*}
with joint posterior distribution given as:
\begin{gather}
\pi(\beta,\theta,t|y) \propto \pi_1(\theta)\pi_2(\beta|\theta)\pi_3(t|\beta,\theta)f(y|t)
\label{posterior}
\end{gather}
where $f(y|t) = \prod_{i=1}^{n}{f(y_i|t_i)}$ is the likelihood function. Here $\Sigma(\theta)$ is the covariance matrix with diagonal elements equal to $\sigma^2+\tau^2$ and off-diagonal elements of $\sigma^2\rho(||s_i-s_j||;\phi,\kappa)$ where $\rho$ is the Mat\'ern correlation function. We consider \emph{Exponential(0.5)} priors for $\sigma$ and $\tau$, and a \emph{Gamma(3,35)} prior for $\phi$.



There are a number of R packages available for posterior estimation of the BGLGM. The \textbf{geostatsp} package by \cite{geostatsp} uses INLA to approximate the marginal posterior distributions, while the recent \textbf{PrevMap} package \citep{PrevMap} uses an MCMC method to generate joint posterior draws.  In this paper, our focus will be on using MCMC methods to generate joint posterior samples of BGLGM. However, \textbf{PrevMap} performed poorly with this dataset when the number of data points was very small, and a bespoke MCMC algorithm was developed as a result.

As reparamterization and standardization help reduce correlation between variables, they will be play an important role in improving the mixing and convergence of MCMC algorithms. The transformations applied to all the model parameters in (\ref{posterior}) follow the recommendations of \cite{transform}. 

Let $\Lambda(t)$ be a diagonal matrix with elements $-\partial^2/\partial t_i^2~ \log f(y_i|t_i)$ for $i=1,...,n$, and denote $\hat{t}_i = \argmax f(y_i|t_i)$. Assuming a prior $N(\mu,\Omega)$ for $\beta$, let $\tilde{\Sigma} = (\Sigma^{-1} + \Lambda(\hat{t}))^{-1}$ and $\tilde{\Omega}=(\Omega^{-1} + X^T(\Sigma^{-1} - \Sigma^{-1}\tilde{\Sigma}\Sigma^{-1})X)^{-1}$. Then by factorizing the posterior distribution in (\ref{posterior}) into two parts:
$\pi(\beta,\theta,t|y) \propto  \pi_1(\theta) f(t,\beta|\theta,y)$, we will be able to simplify the second factor $f(t,\beta|\theta,y)$ as following:
\begin{align}
\begin{split}
\log f(t,\beta|\theta,y) & \approx -0.5(t-\hat{t})^T\Lambda(\hat{t})(t-\hat{t}) - 0.5(t-X\beta)^T\Sigma^{-1}(t-X\beta) - 0.5(\beta-\mu)^T\Omega^{-1}(\beta-\mu)\\
& = - 0.5(t- \tilde{\Sigma}(\Lambda(\hat{t})\hat{t} + \Sigma^{-1}X\beta))^T\tilde{\Sigma}^{-1}(t- \tilde{\Sigma}(\Lambda(\hat{t})\hat{t} + \Sigma^{-1}X\beta))\\
& ~~~ -0.5(\beta-\tilde{\Omega}(X^T\Sigma^{-1}\tilde{\Sigma}\Lambda(\hat{t})\hat{t} + \Omega^{-1}\mu))^T\tilde{\Omega}^{-1}(\beta-\tilde{\Omega}(X^T\Sigma^{-1}\tilde{\Sigma}\Lambda(\hat{t})\hat{t} + \Omega^{-1}\mu))
\end{split}
\label{simplify_post}
\end{align}
where the first expression $-0.5(t-\hat{t})^T\Lambda(\hat{t})(t-\hat{t})$ is derived from the Taylor expansion of $\log f(y|t)$ around $\hat{t}$. From equation (\ref{simplify_post}), we can simply use the transformations:
\begin{align}
\boldsymbol{\tilde{t}} & = (\tilde{\Sigma}^{1/2})^{-1}(\boldsymbol{t} - \tilde{\Sigma}(\Lambda(\hat{t})\hat{t} + \Sigma^{-1}X\beta))\label{t_trans}\\
\boldsymbol{\tilde{\beta}} & = (\tilde{\Omega}^{1/2})^{-1}(\boldsymbol{\beta}-\tilde{\Omega}(X^T\Sigma^{-1}\tilde{\Sigma}\Lambda(\hat{t})\hat{t} + \Omega^{-1}\mu)) \label{beta_trans}
\end{align}
where $\tilde{t}_1,...,\tilde{t}_n$ and $\tilde{\beta}_1,...,\tilde{\beta}_p$ are now approximately uncorrelated with mean zero and variance one. These parameters are also uncorrelated with $\theta$ and hence there will be no posterior dependence between $\tilde{t}, \tilde{\beta},$ and $\theta$. However, according to \cite{transform} and \cite{PrevMap}, there is posterior dependence within the parameters of $\theta^T = (\theta_1,\theta_2,\theta_3) = (\sigma^2, \phi,\tau)$, and hence a reparameterization is proposed as following:
$$\boldsymbol{{\tilde{\theta}}} = (\tilde{\theta}_1,\tilde{\theta}_2,\tilde{\theta}_3) = (\log\sigma, \log \sigma^2/\phi^{2\kappa}, \log \tau^2)$$

In general, using these transformations will help facilitate the choice of proposal densities as well as reducing the correlation between variables that will significantly improve mixing and convergence of the MCMC algorithm.

Using these reparameterizations, we have implemented a Metropolis-Hastings-within-Gibbs sampling method that updates each blocks of $\tilde{\theta}, \tilde{\beta},$ and $\tilde{t}$ at a time. However, for high-dimensional parameters, it is more suitable to use the Langevin-Hastings algorithm as they will have much faster convergence rates \citep{langevin,roberts_tweedie,moller}. For our model and data, $\tilde{t}$ has the highest dimension, hence we will use Langevin-Hastings algorithm to update $\tilde{t}$. For the remaining blocks we will use the Random-Walk Metropolis Hastings (RWMH) algorithm. The summary of the steps used to run the MCMC algorithm is shown in the diagram below.


\begin{algorithm}[h]
Initialize $\theta, \beta,$ and $t$\\
Transform to $\tilde{\theta}, \tilde{\beta},$ and $\tilde{t}$\\
Update $\tilde{\theta}_1, \tilde{\theta}_2$ and $\tilde{\theta}_3$ using a RWMH, each with standard deviation $s_i$ calculated iteratively as: $$s_i = s_{i-1} + c_1 i^{-c_2}(\alpha_i-0.45)$$ where $c_1>0$ and $c_2 \in (0,1]$ are constants, and $\alpha_i$ is the acceptance probability up to $i-th$ iteration with optimal acceptance probability of $0.45$.\\
Update $\tilde{\beta}$ using a RWMH\\
Update $\tilde{t}$ with a Langevin-Hastings algorithm, i.e. $\tilde{t} \textprime \sim MVN(\tilde{t} + 0.5 h \nabla \log \pi(\tilde{t}), h I)$ where $h$ is recommended to be $1.65^2/n^{1/3}$.\\
Repeat steps 3-5 until the desired number of samples are collected.\\
Transform samples of $\tilde{\theta}, \tilde{\beta},$ and $\tilde{t}$ back to $\theta, \beta,$ and $t$.
\caption{MCMC algorithm}
\label{mcmc}
\end{algorithm}

\subsection{Prediction \& Assessment}
After running our MCMC algorithm on the BGLGM, we will combine the posterior samples for each parameter to generate posterior distributions for hardwood probabilities at each of the 62 validation locations. We will then emphasize on assessing the predictions from the number of hardwood counts rather than proportions, since the observed proportions are often 0 or 1, while predictions are $0<p<1$. Below we describe the various assessments we have considered:
\begin{enumerate}
\item \textit{Coverage Probability}: 
For each of the 62 validation points, we generate posterior samples of hardwood counts from the corresponding posterior probability samples, then examine whether the true hardwood count is inside the (say) 95\% posterior interval. The coverage probability will be the proportion of 62 points that are inside their posterior intervals, i.e.: $$\# (true ~hardwood ~count ~\in~ posterior ~interval ~of ~hardwood~ counts)/62$$
\item \textit{RMSE (root mean squared error)}:
We will also compare RMSE of hardwood probabilities from both BGLGM and GLM (Logistic Regression), calculated as:
$$ RMSE = \sqrt{\frac{1}{62}\sum_{j=1}^{62}{(\hat{p}_j - p_j)^2} }$$ where $p_j$ is the true proportion of hardwoods in plot $i$ (often 0 or 1) and $\hat{p}_j $ is the predicted hardwood probability in GLM and posterior mean in BGLGM. 
\item \textit{Total hardwood count distribution}:
We also consider the distribution of the total number of hardwoods in \textit{all} 62 validation sites and examine whether the true total hardwood counts is covered within the 95\% posterior interval. Unlike the posterior distributions of hardwood counts in each of the 62 plot, the total count has a reasonably symmetric distribution. In addition, we have compared this to the corresponding distribution generated from GLM via bootstrapping. 

\end{enumerate}

\section{Results}\label{results}
For the main analysis we have ran the MCMC algorithm for 2,000,000 iterations with 1,000,000 burnin and 100 thinning. Runs consist of fitting 100, 25, and 10 sites as training data, both via random and stratified sampling, with predictions made for the 62 validation data. We have repeated this procedure for five different simulations by randomly choosing five different validation sets of size 62.

%

\begin{figure}[!h]
\centering
\begin{subfigure}{.45\textwidth}
  \centering
  \includegraphics[width=1\linewidth]{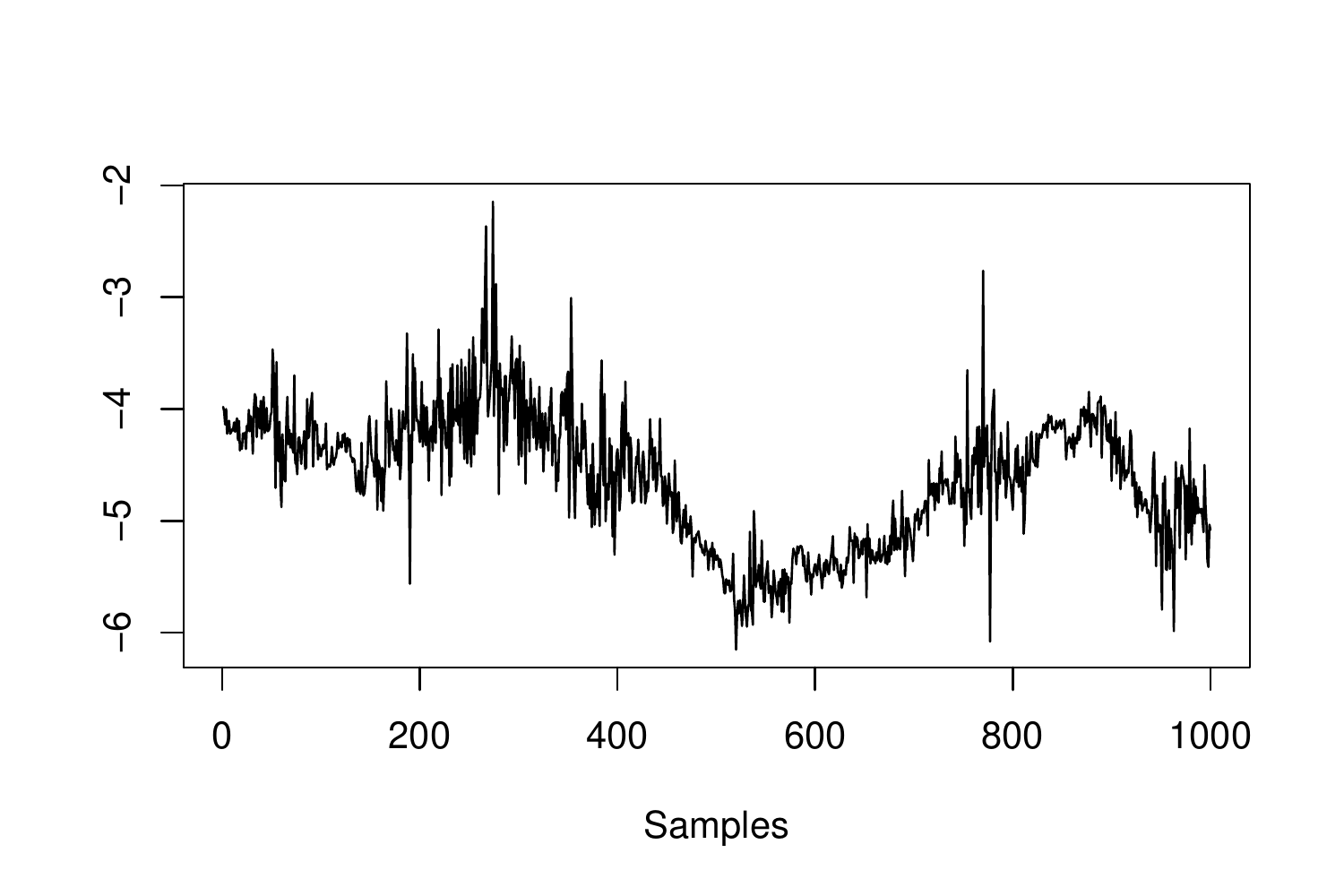}
  \caption{PrevMap, Intercept}
\end{subfigure}
\begin{subfigure}{.45\textwidth}
  \centering
  \includegraphics[width=1\linewidth]{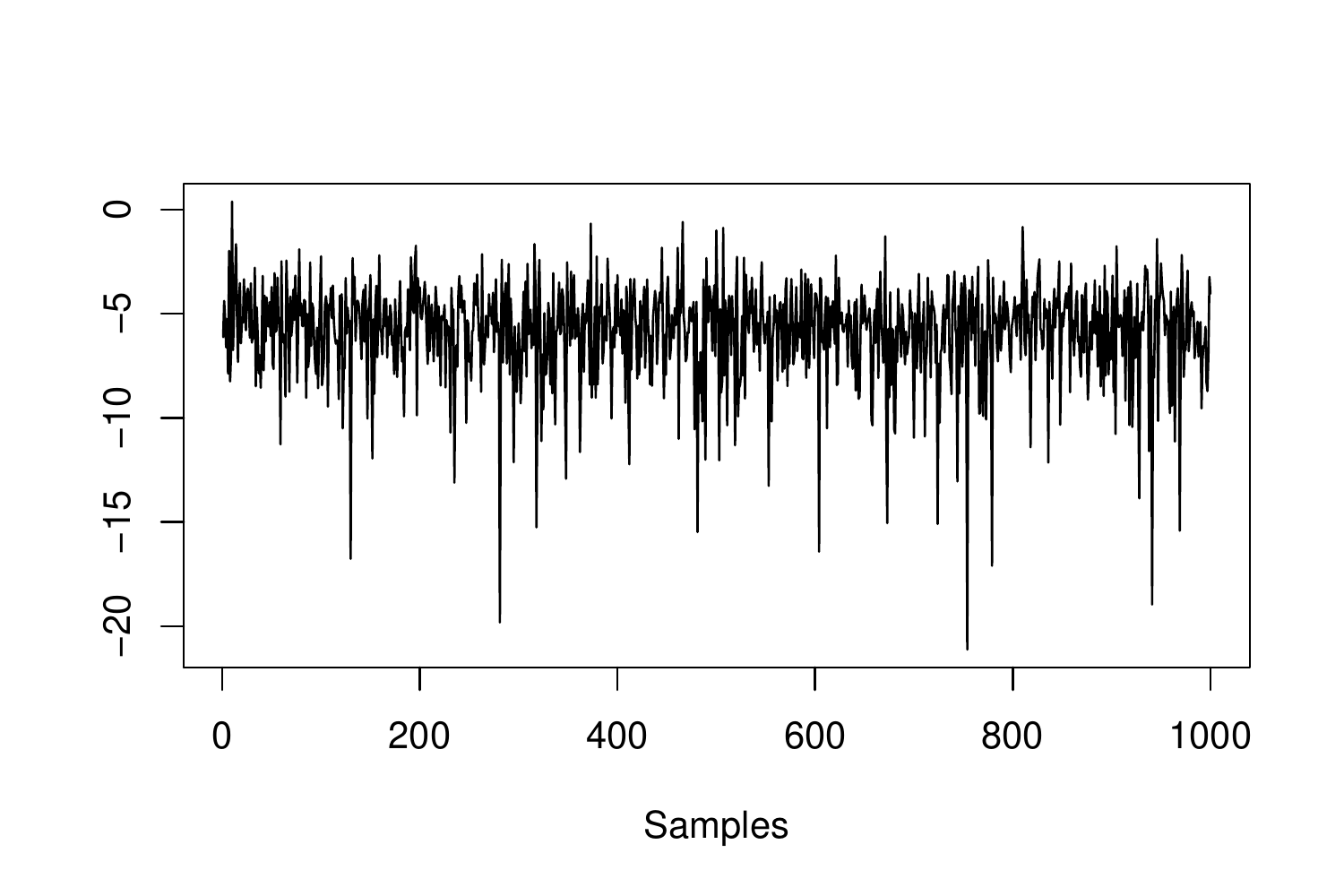}
  \caption{Bespoke MCMC, Intercept}
\end{subfigure}
\begin{subfigure}{.45\textwidth}
  \centering
  \includegraphics[width=1\linewidth]{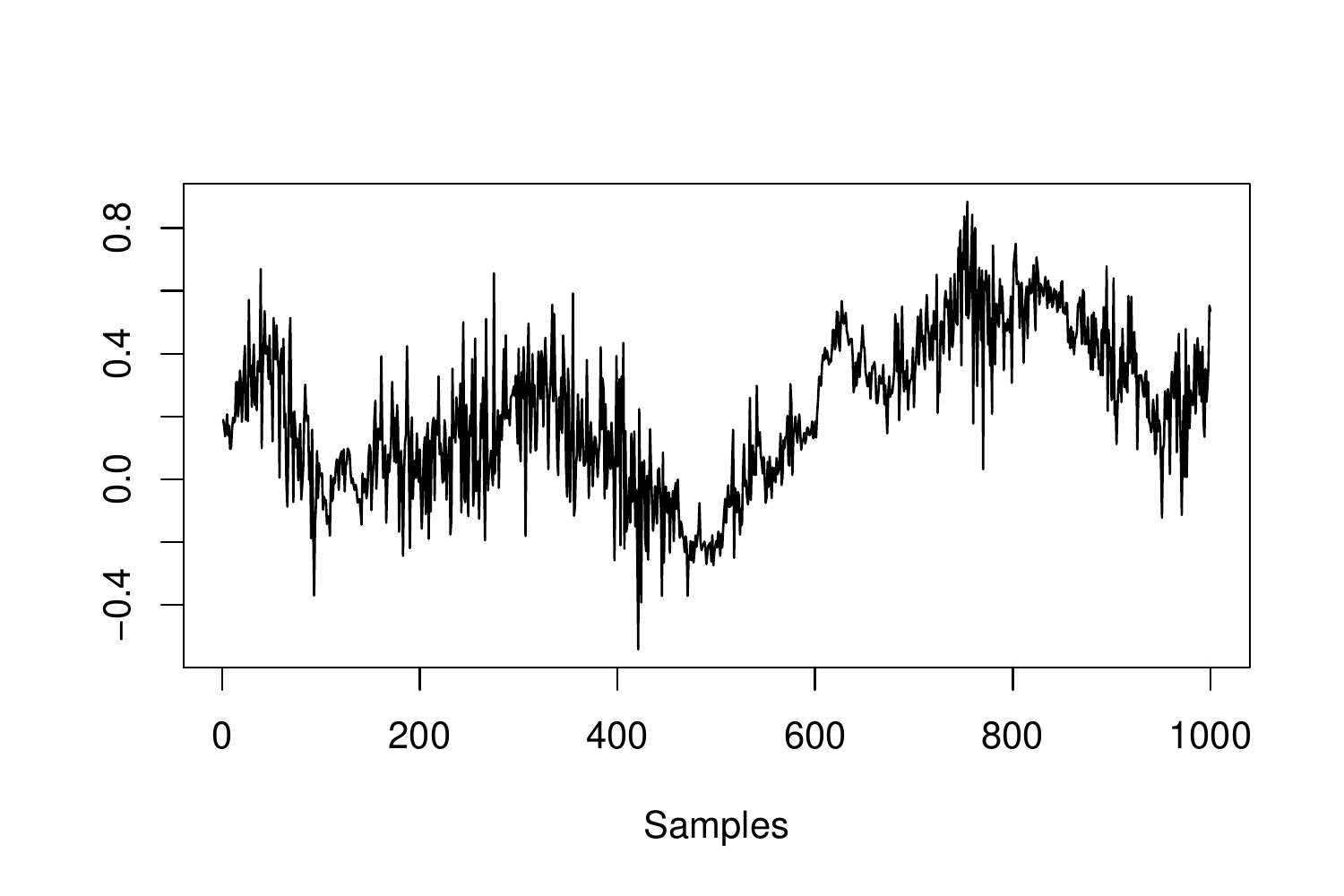}
  \caption{PrevMap, elevation}
\end{subfigure}
\begin{subfigure}{.45\textwidth}
  \centering
  \includegraphics[width=1\linewidth]{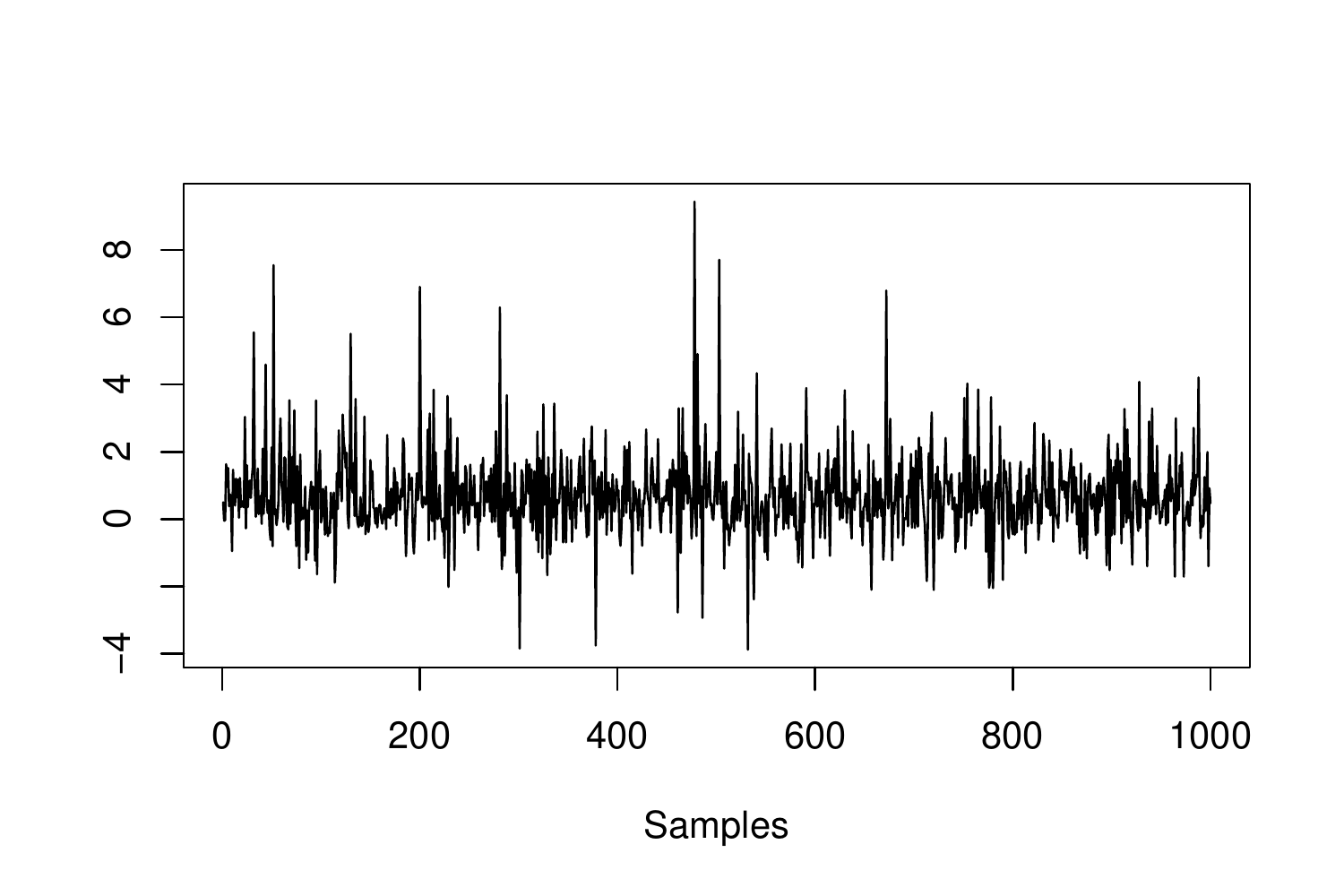}
  \caption{Bespoke MCMC, elevation}
\end{subfigure}
\caption{Comparing trace plots of $\beta_0$ and $\beta_1$ from the bespoke MCMC implementation and the PrevMap package.}
\label{prevmap_our1}
\end{figure}


\subsection{MCMC Convergence and Mixing}

Figure \ref{prevmap_our1} is showing the trace plots of posterior samples for $\beta$ generated from the bespoke MCMC implementation from section \ref{bespoke_mcmc} versus PrevMap package, using 10 training sites. The bespoke MCMC mixes well and has converged, while PrevMap trace plots have not converged although they have been ran for the same number of iterations. Different tuning parameters for running PrevMap were considered without the chains being improved. Thus we will be using the bespoke MCMC implementation for the rest of the analysis in this paper.




\begin{figure}[!h]
\centering
\begin{subfigure}{.45\textwidth}
  \centering
  \includegraphics[width=1\linewidth]{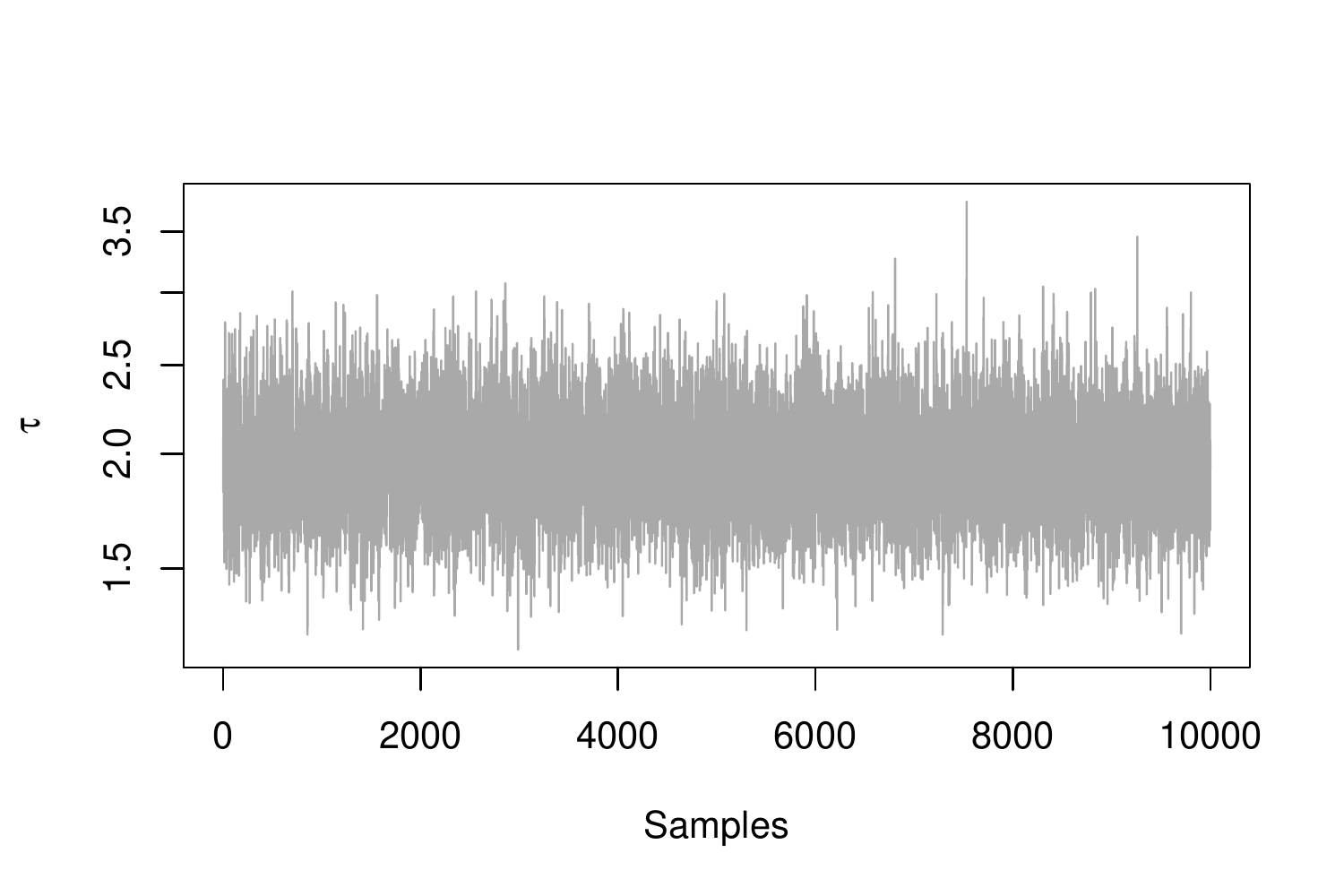}
  \caption{100 training sites --- $\tau$}
  \label{trace_tau_100}
\end{subfigure}
\begin{subfigure}{.45\textwidth}
  \centering
  \includegraphics[width=1\linewidth]{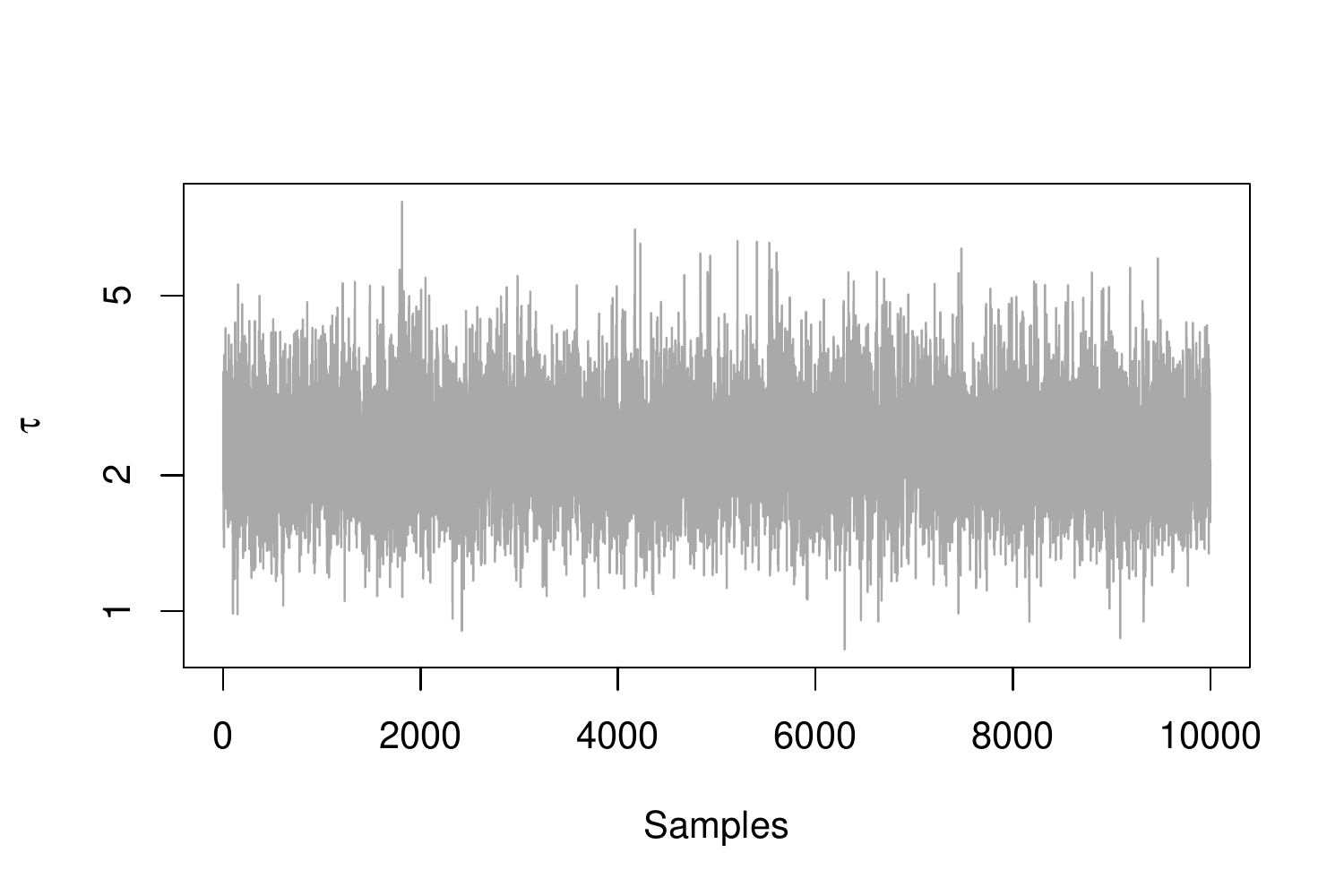}
  \caption{25 training sites --- $\tau$}
  \label{trace_tau_25}
\end{subfigure}
\begin{subfigure}{.45\textwidth}
  \centering
  \includegraphics[width=1\linewidth]{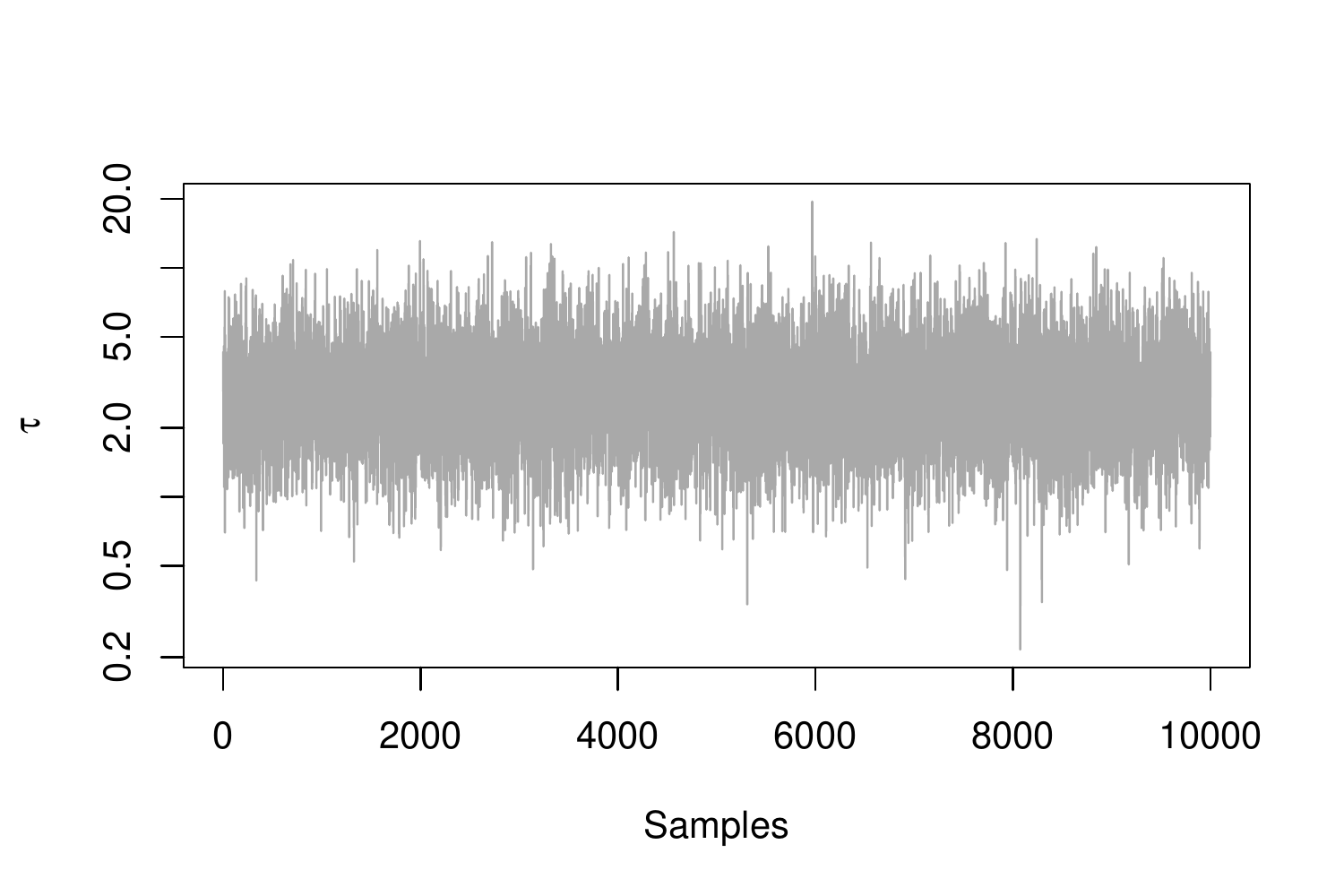}
  \caption{10 training sites --- $\tau$}
  \label{trace_tau_10}
\end{subfigure}
\caption{Trace plots of 10,000 MCMC posterior samples for $\tau$ (simulation 1).}
\label{trace_tau}
\end{figure}

Figures \ref{trace_tau_100}, \ref{trace_tau_25}, and \ref{trace_tau_10} are showing the MCMC trace plots for only the $\tau$ parameter with 100, 25, and 10 data fitted to the model respectively. 
All trace plots show that the MCMC is mixing well and thus, the chains have converged. In addition, the trace plots show larger variability with less training data fitted. The remaining trace plots for other parameters as well as other simulations are included in the appendix.

\begin{table*}[!h]
\footnotesize
\centering
\caption{Comparison of posterior mean, 2.5 \%, and 97.5 \% quantiles of model parameters, for different sizes of training data. These results are from only the first of five training samples.}
\label{quantiles}
\begin{tabular}{cccccc}
\hline \\
Parameters & \# of training & Mean & $2.5\%$ quantile & $97.5 \%$ quantile  \\ \\
\hline
  & 100 & -3.47  & -4.33 & -2.65    \\ 
Intercept - $\beta_0$ & 25 & -3.54 & -5.67 &  -1.66  \\ 
 & 10 & -2.37 & -6.38 & 1.31  \\ \\

 & 100 & 0.53  & 0.07 & 0.99   \\ 
Elevation - $\beta_1$ & 25 & 0.12 & -1.03  & 1.13  \\ 
 & 10 & 2.16 & -0.96 &  6.38  \\ \\

 & 100 & 2.89  & 1.19 & 4.87    \\ 
SkyF$<$0.3 - $\beta_2$ & 25 & 2.09 & -0.50 & 5.26  \\ 
& 10 & 3.38 & -1.39 & 10.42  \\ \\

 & 100 & 2.61 & 2.10  & 3.17   \\ 
SkyF$>$0.3 - $\beta_3$ & 25 & 3.02 & 1.86 & 4.44  \\ 
 & 10 & 4.20 & 1.85 & 7.70  \\ \\

 & 100 & 0.04  & 0.02   & 0.11    \\ 
Spatial sd - ${\sigma}$ & 25 & 0.04 & 0.02   &  0.12  \\ 
& 10 & 0.06 & 0.02  &  0.17  \\ \\

 & 100 & 1.98  & 1.52   & 2.55   \\ 
Indep. sd - ${\tau}$ & 25 & 2.38 & 1.36 & 4.05  \\ 
 & 10 & 3.09 & 1.04 & 7.23 \\ \\

 & 100 & 104.94  & 21.89 & 255.14    \\ 
Range(km) - ${\phi}$ & 25 & 105.42 & 22.00 & 252.93  \\ 
 & 10 & 105.06 & 21.30 & 255.27  \\ \\

 \hline
\end{tabular}
\end{table*}

For quantitively verifying this variability between different training data size, we have compared the numerical values of posterior mean, 2.5 \%, and 97.5 \% quantiles of all model parameters in Table \ref{quantiles}. While the posterior means remain almost unchanged, the 95\% posterior intervals for each model parameter (except $\phi$), become wider with less training data fitted to the model, indicating more uncertainty in parameter estimation. 

\subsection{Parameter posteriors \& spatial surfaces}

\begin{figure}[!h]
\centering
\begin{subfigure}{.47\textwidth}
  \centering
  \includegraphics[width=1\linewidth]{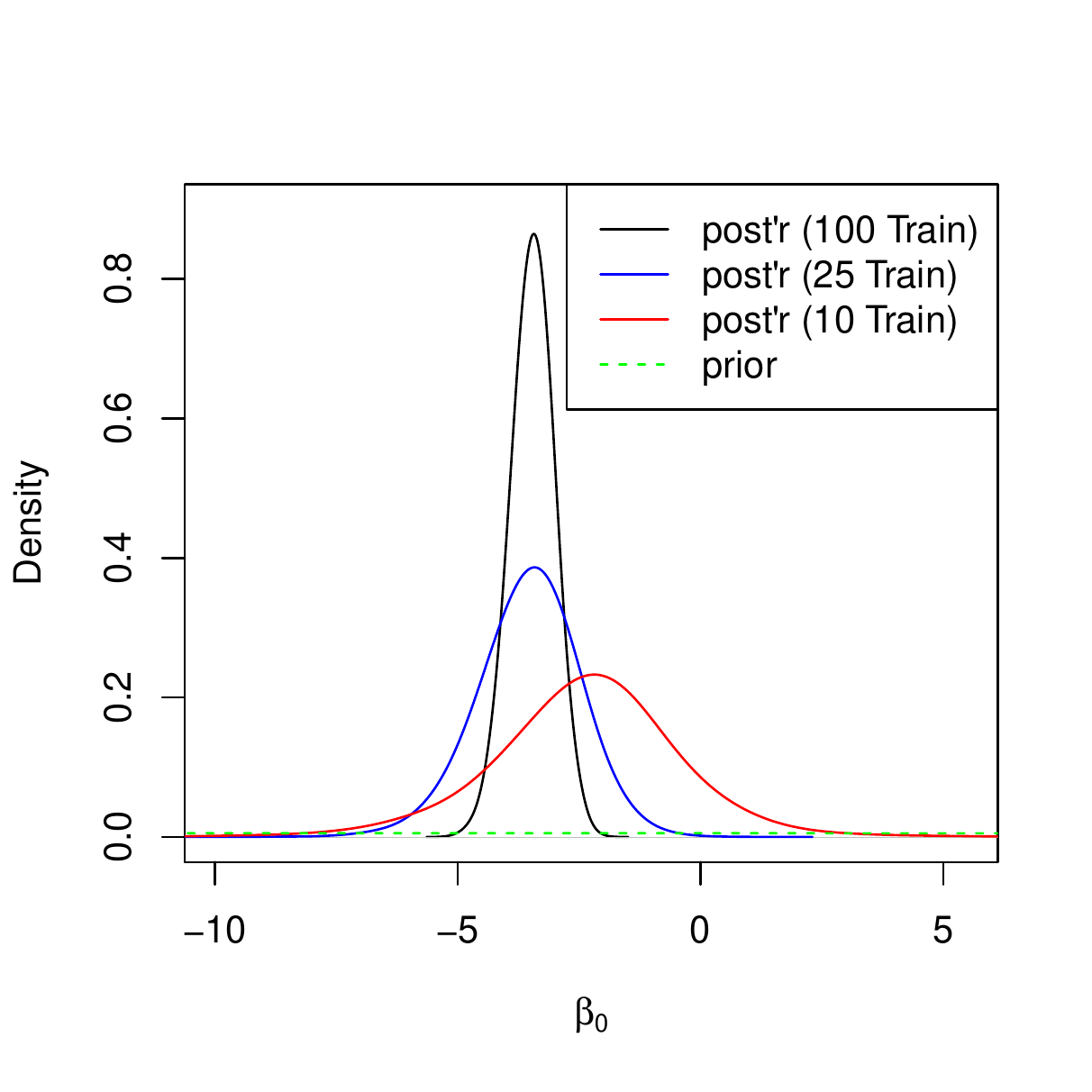}
  \caption{$\beta_0$ - Intercept}
  \label{post_beta0}
\end{subfigure}
\begin{subfigure}{.47\textwidth}
  \centering
  \includegraphics[width=1\linewidth]{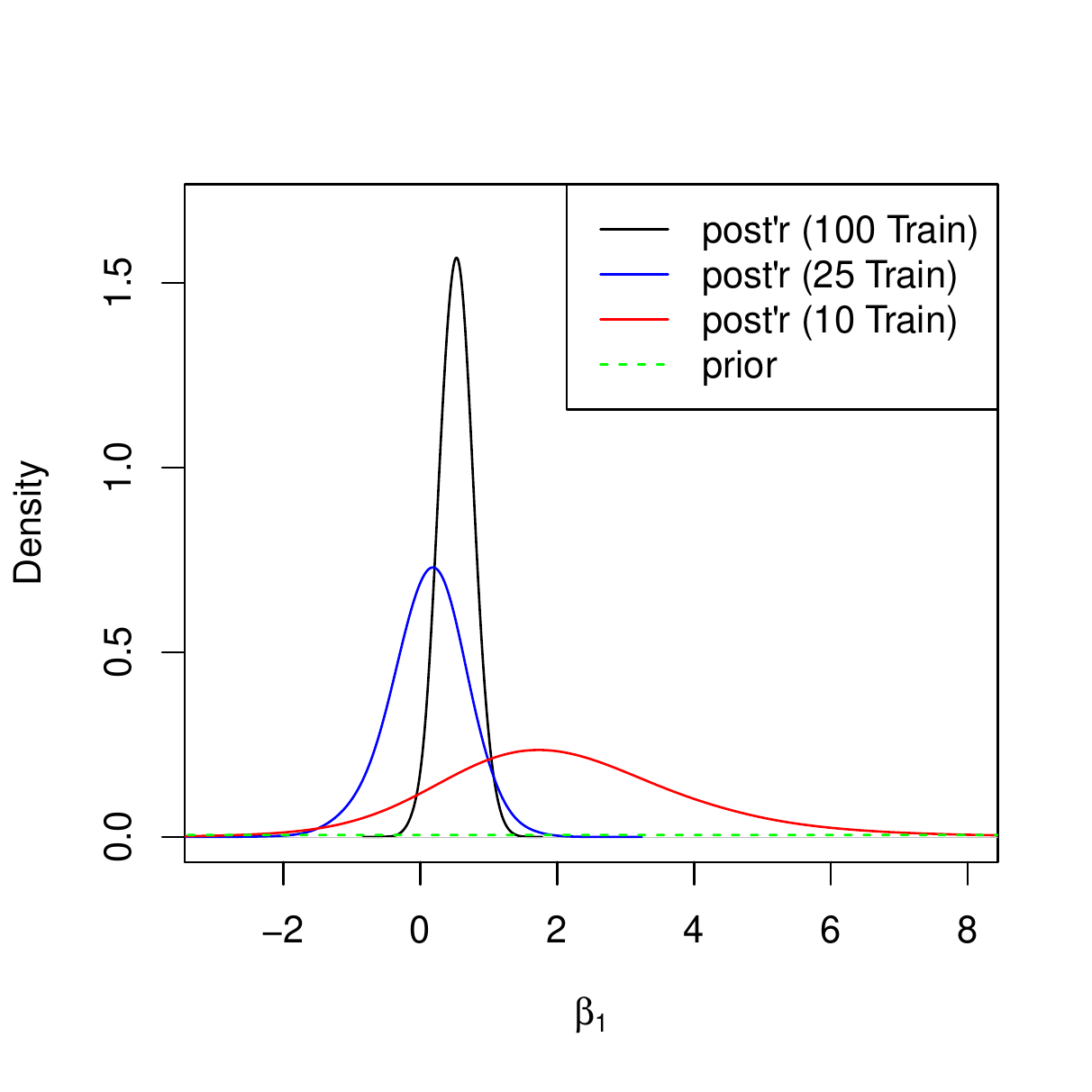}
  \caption{$\beta_{1}$ - elevation}
  \label{post_beta1}
\end{subfigure}
\begin{subfigure}{.47\textwidth}
  \centering
  \includegraphics[width=1\linewidth]{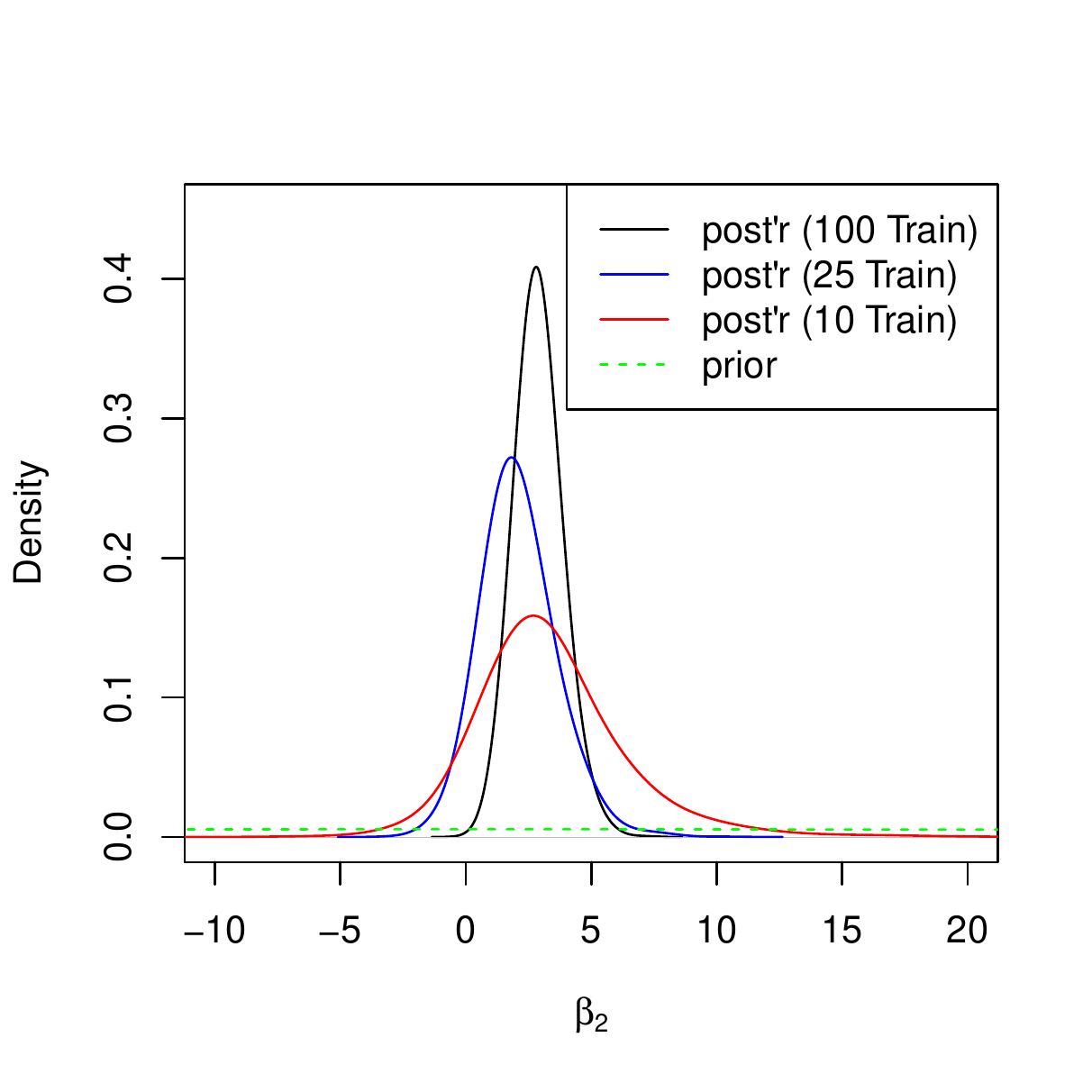}
  \caption{$\beta_{2}$ - SkyForest$^{\text{TM}}$ vegetation below $0.3$}
  \label{post_beta2}
\end{subfigure}
\begin{subfigure}{.47\textwidth}
  \centering
  \includegraphics[width=1\linewidth]{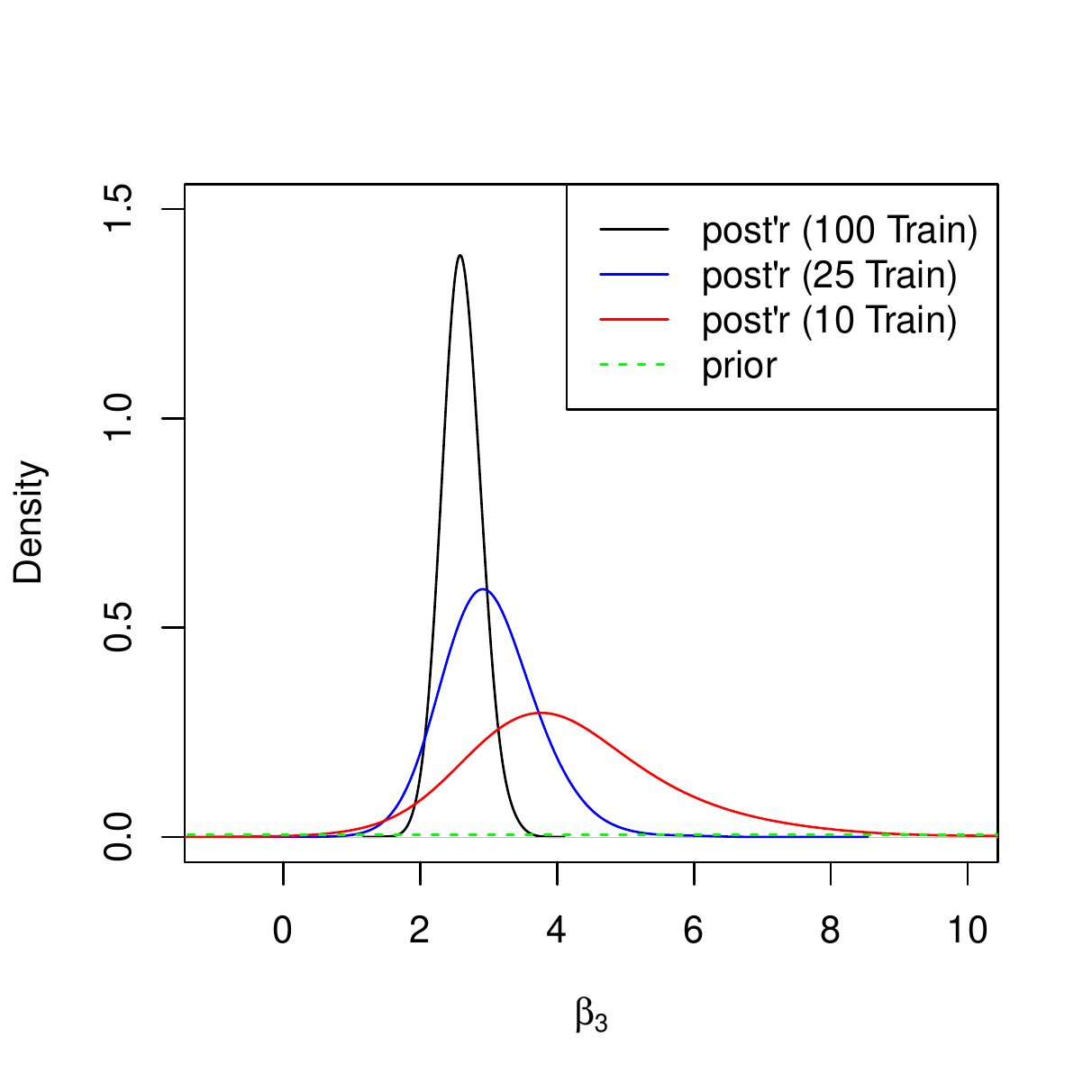}
  \caption{$\beta_{3}$ - SkyForest$^{\text{TM}}$ vegetation above $0.3$}
  \label{post_beta3}
\end{subfigure}
\caption{Prior and posterior distributions of parameters from the first simulation.}
\label{post_betas}
\end{figure}

\begin{figure}[!h]
\centering
\begin{subfigure}{.47\textwidth}
  \centering
  \includegraphics[width=1\linewidth]{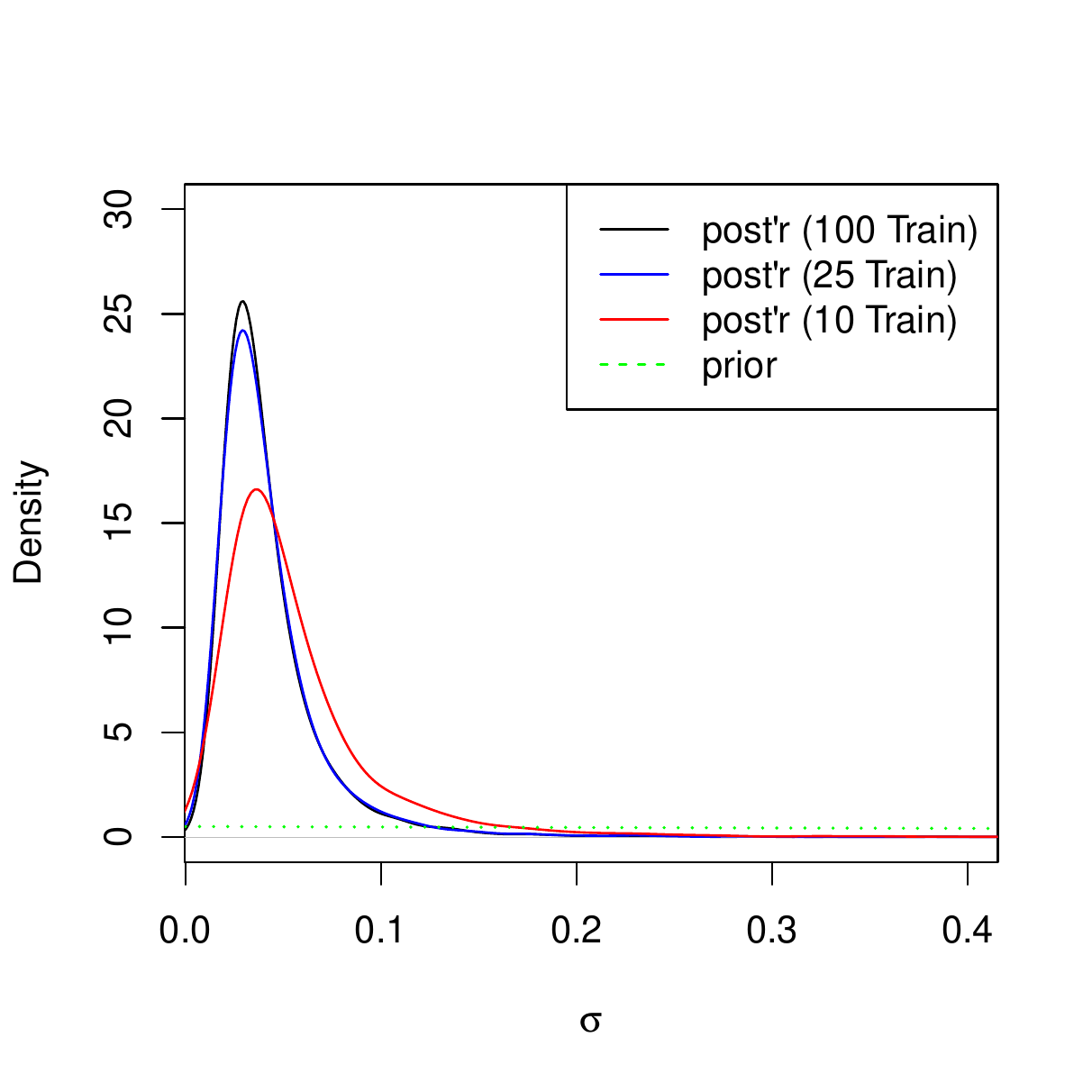}
  \caption{$\sigma$ - spatial sd}
  \label{post_sigma}
\end{subfigure}
\begin{subfigure}{.47\textwidth}
  \centering
  \includegraphics[width=1\linewidth]{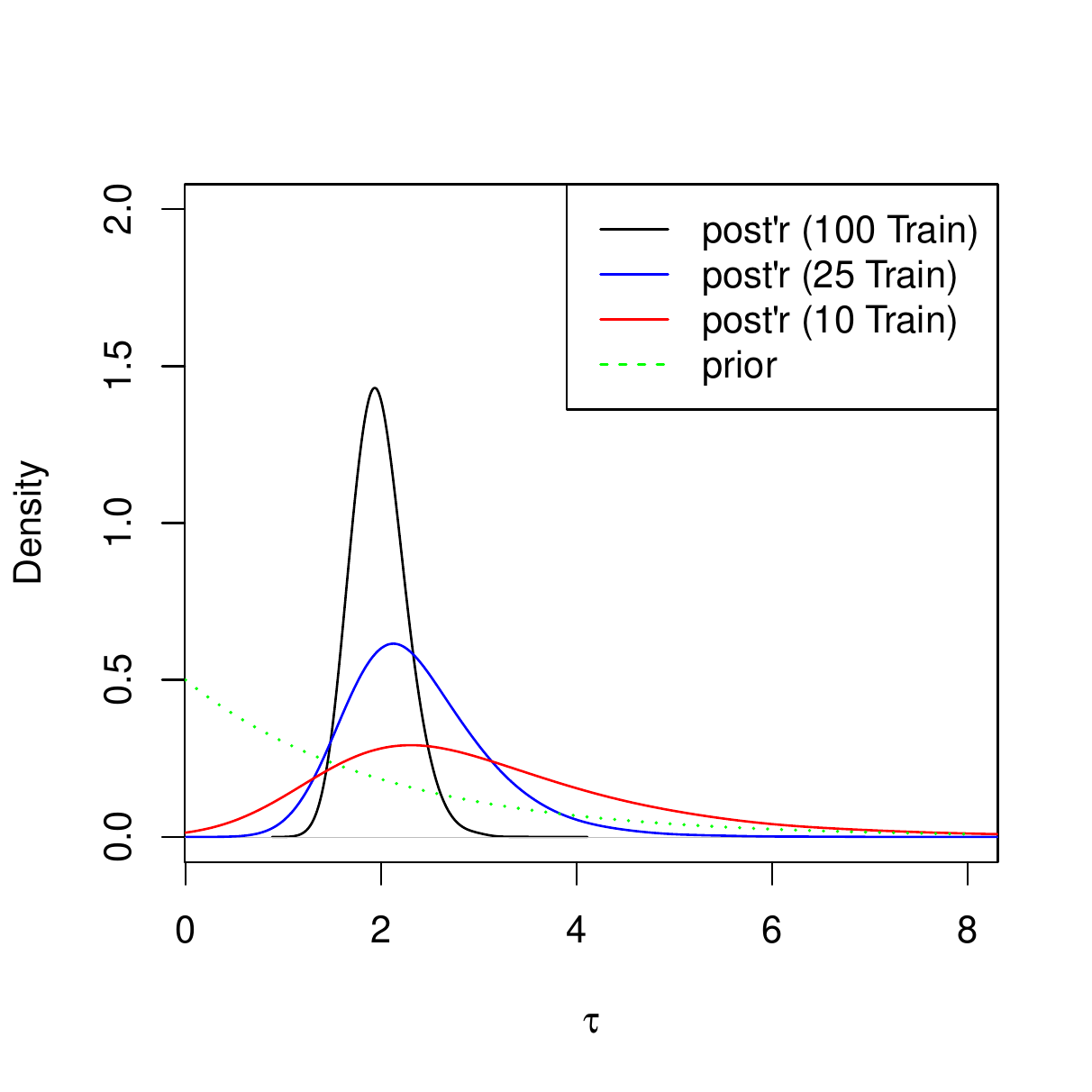}
  \caption{$\tau$ - indep sd}
  \label{post_tau}
\end{subfigure}
\begin{subfigure}{.47\textwidth}
  \centering
  \includegraphics[width=1\linewidth]{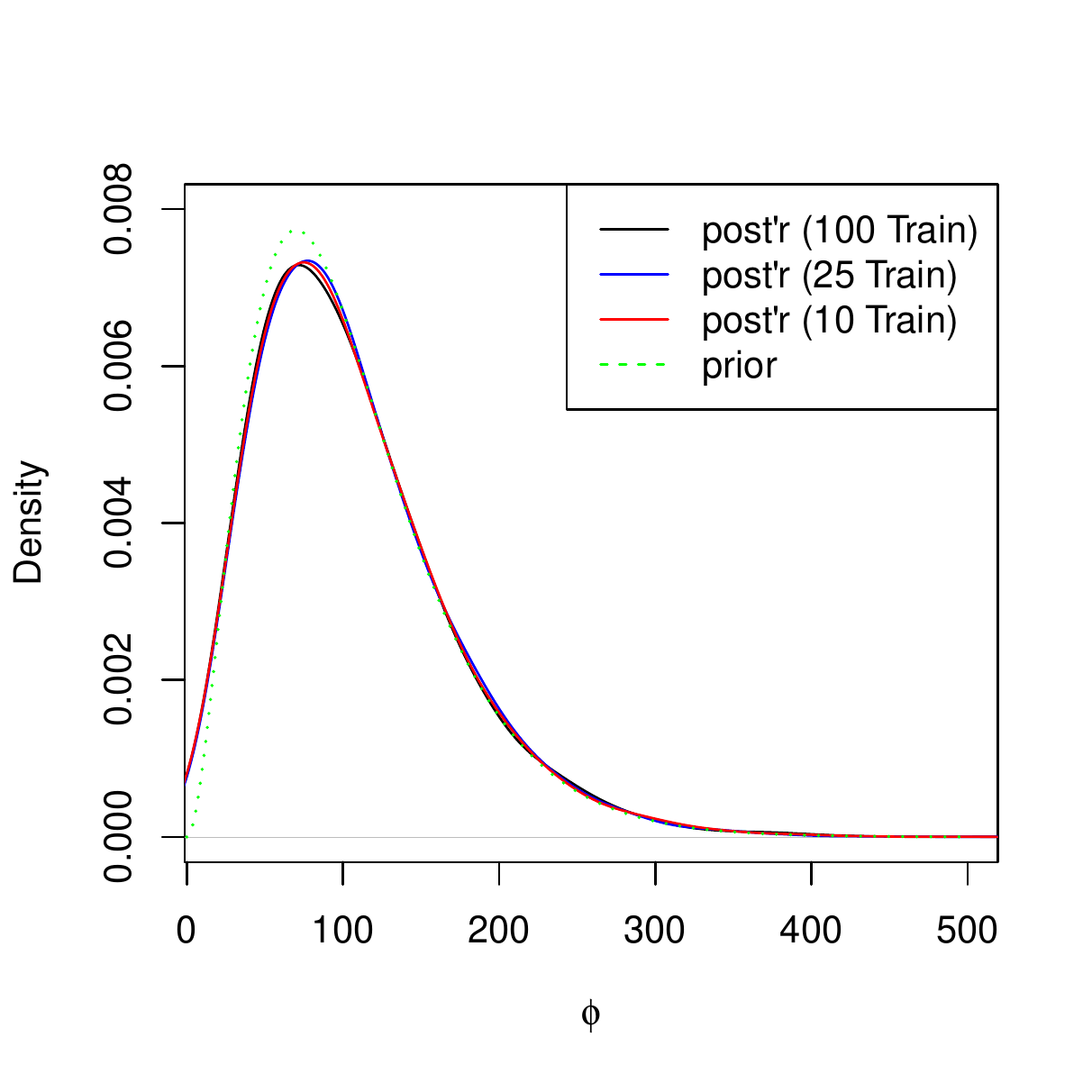}
  \caption{$\phi$ (in km) - range}
  \label{post_phi}
\end{subfigure}
\caption{Priors and posteriors from the first simulation.}
\label{post_thetas}
\end{figure}

The prior and posterior densities of model parameters from the first simulation are shown in Figures \ref{post_betas} and \ref{post_thetas}. From these figures we can ascertain that with fewer training data, posterior densities become wider and hence result in more uncertainty of predictions. The posterior distributions of $\sigma$ suggest small spatial random effects for this dataset, as they have modes concentrated at smaller values. Posterior densities of $\phi$ are all similar and remain unchanged for different training data, as small $\sigma$ causes weak spatial signal which can't identify $\phi$.

One surprising feature of Figure \ref{post_tau}, is the posterior density with 10 training data points does not resemble the prior. Even the smallest training dataset considered provides clear evidence that there is more variation in the observed counts than the covariates predict, which is manifest in the results as $\tau$ has  a posterior distribution concentrated away from zero. There is also evidence that this extra variation is not spatially structured, since $\sigma$ is clearly much smaller than $\tau$. 

The main goal is to predict the composition of trees at unmeasured sites in the forests via simulating posterior samples of $U(g_l)$ for new locations $g_l:$ $l=1,...,L$, conditional on MCMC posteriors $\{U(s_i)+Z(s_i): i=1,...,n\}$. Considering a $100\times100$ grid with $L = 10,000$ cells inside the forests as our new locations, we can simulate $U(g_l)$ using the {\tt RFsimulate} function in the ``RandomFields" package and make predictions for hardwood probabilities $p(g_l)$ for each cell. The RandomFields package has very efficient algorithms for simulating from conditional distributions of spatial processes without using the full variance matrix. Thus assuming we have grid cells $g_1,...,g_L$, we simulate $[U(g_1),...,U(g_L) | Y]$ and independent $Z_1,...,Z_L$ along with the use of other posterior samples to generate $[p(g_1),...,p(g_L) | Y]$.


\begin{figure}[!h]
\centering
\begin{subfigure}{.3\textwidth}
  \centering
  \includegraphics[width=0.88\linewidth]{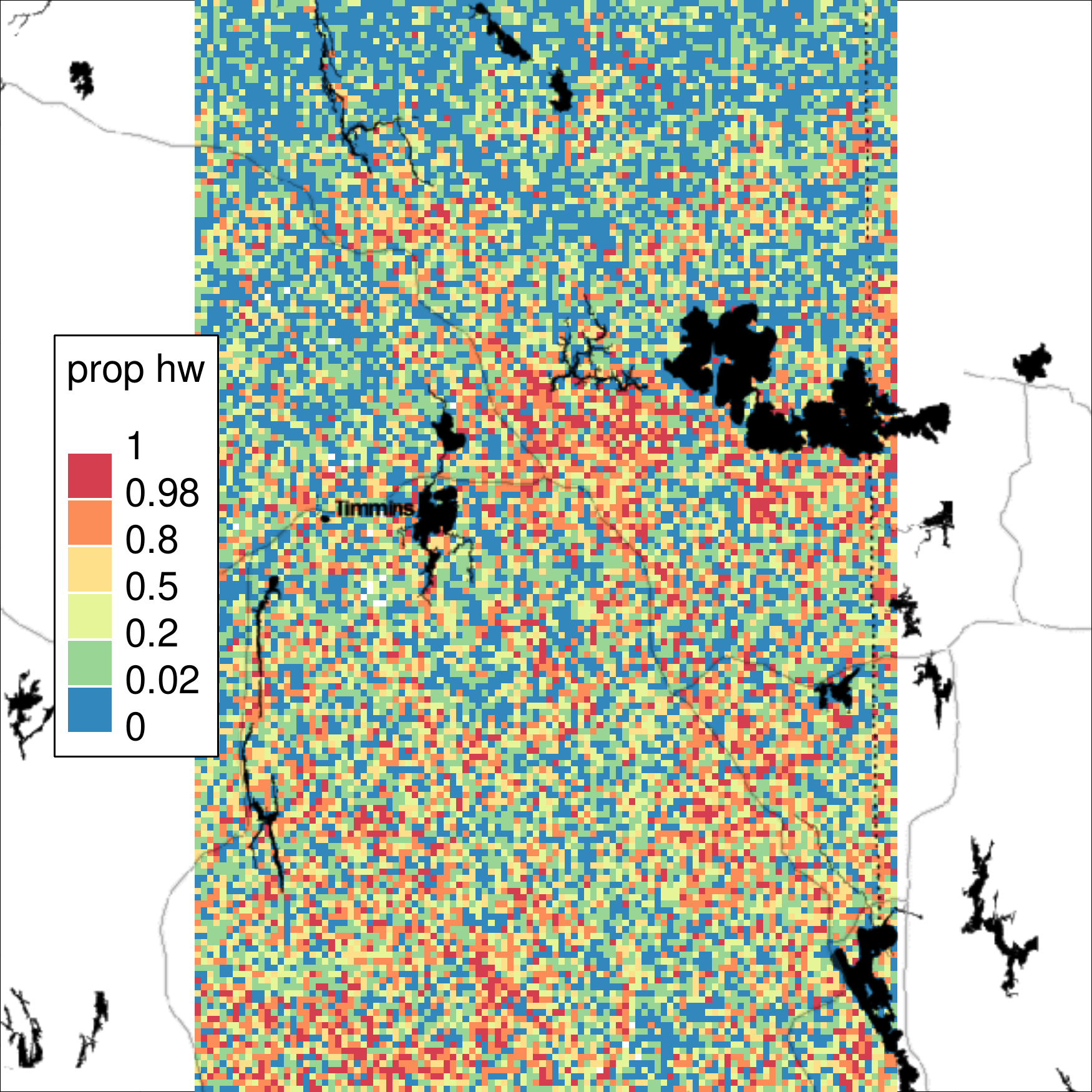}
  \caption{100 train - Sample 1}
  \label{100-1}
\end{subfigure}
\begin{subfigure}{.3\textwidth}
  \centering
  \includegraphics[width=0.88\linewidth]{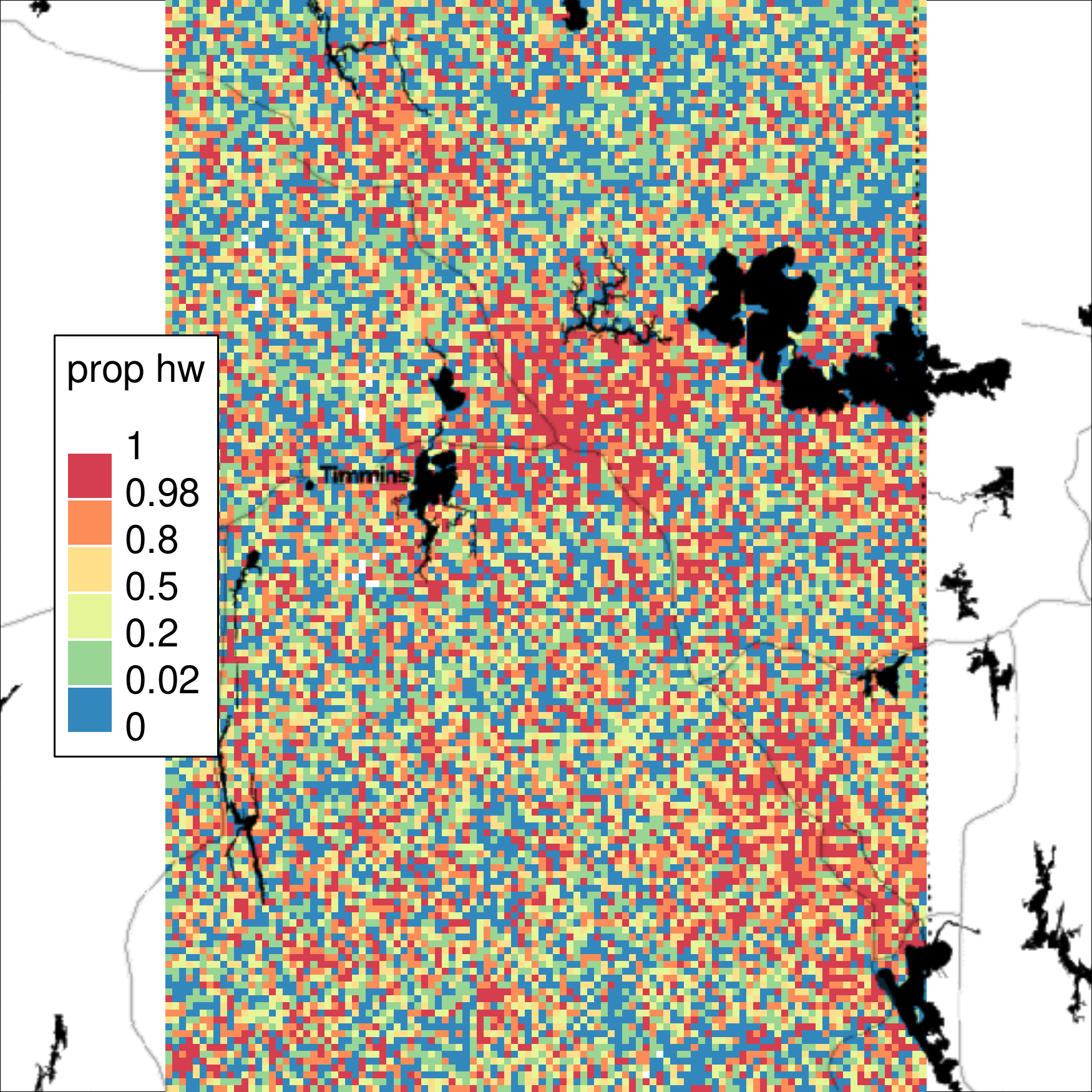}
  \caption{25 train - Sample 1}
  \label{100-2}
\end{subfigure}
\begin{subfigure}{.3\textwidth}
  \centering
  \includegraphics[width=0.88\linewidth]{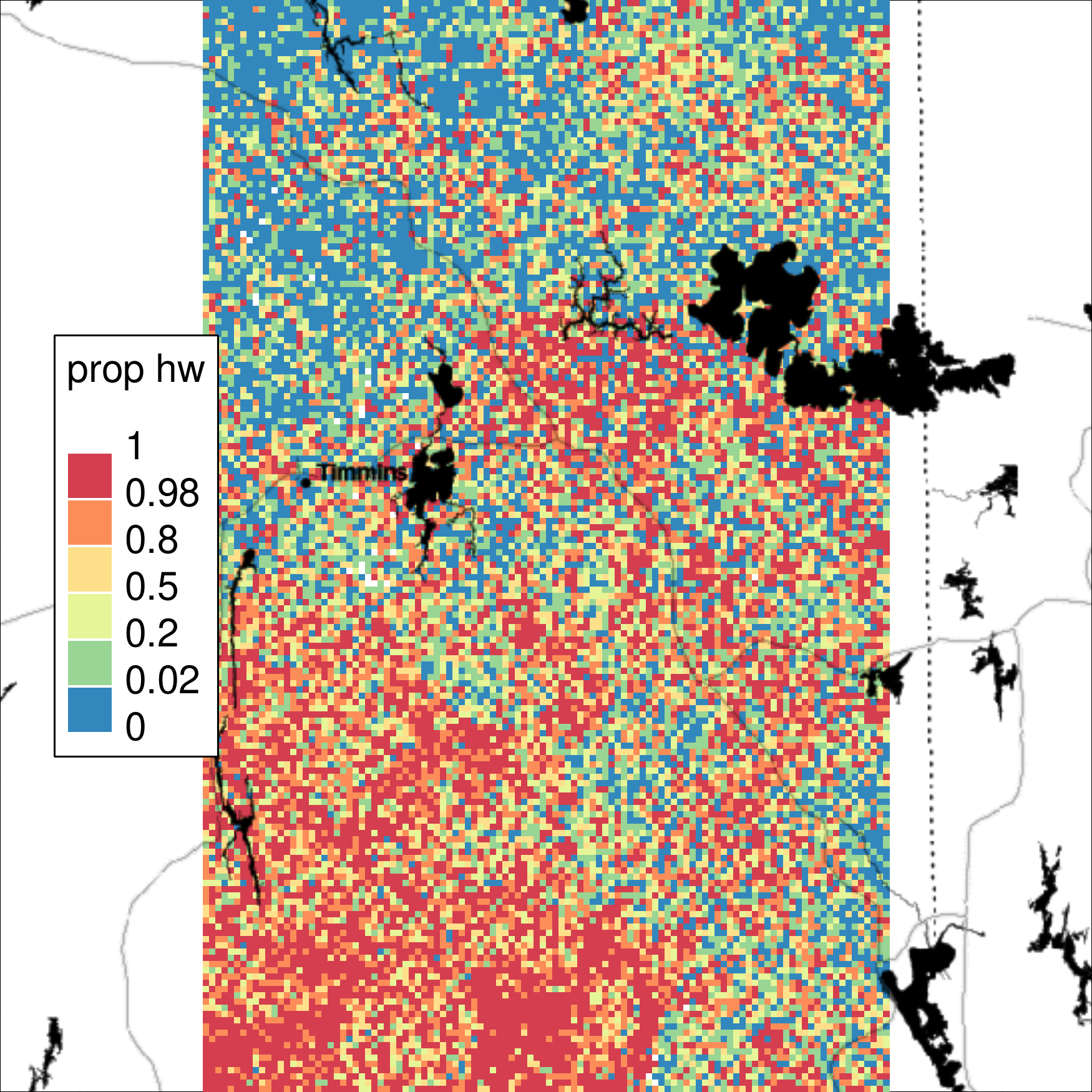}
  \caption{10 train - Sample 1}
  \label{100-3}
\end{subfigure}
\begin{subfigure}{.3\textwidth}
  \centering
  \includegraphics[width=0.88\linewidth]{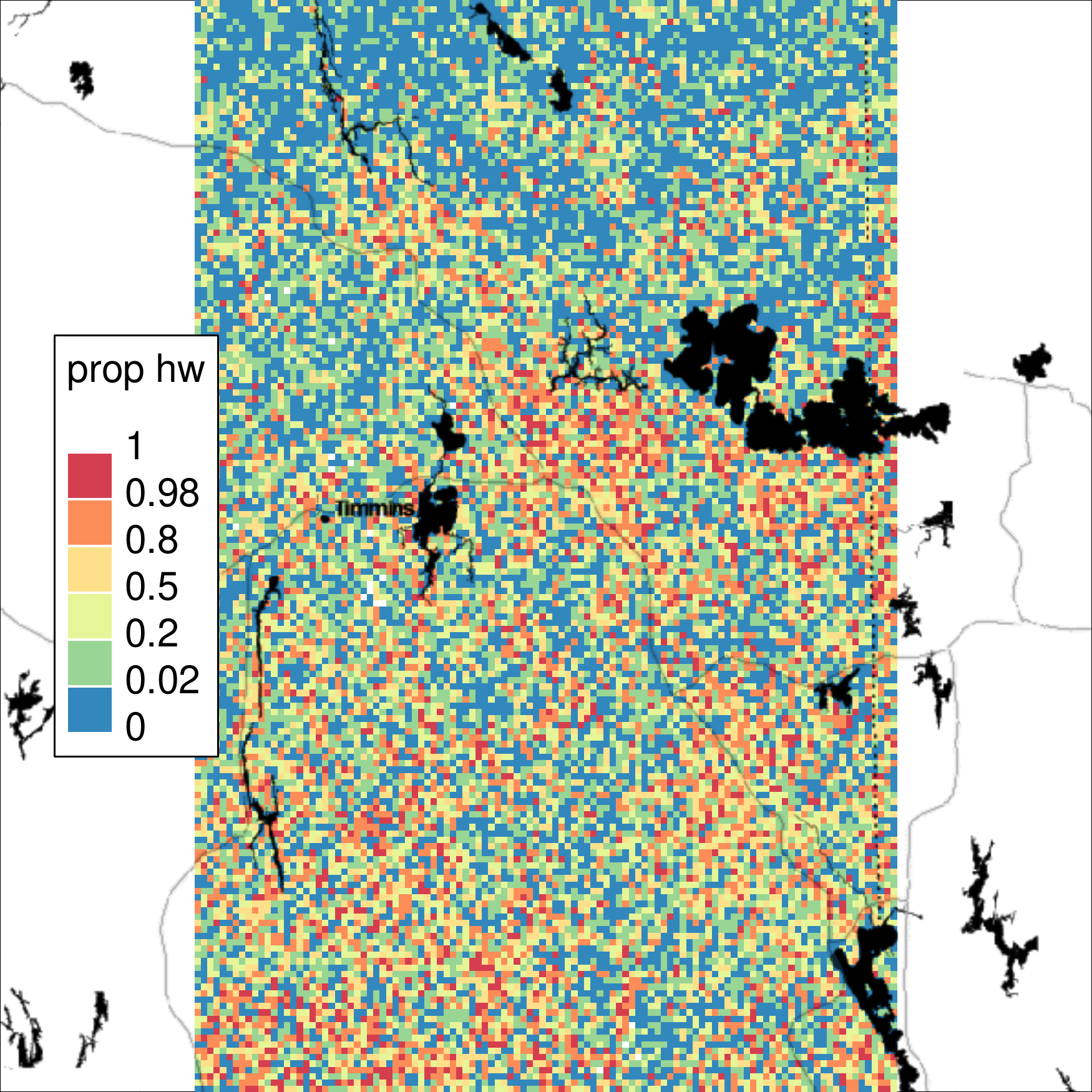}
  \caption{100 train - Sample 2}
  \label{25-1}
\end{subfigure}
\begin{subfigure}{.3\textwidth}
  \centering
  \includegraphics[width=0.88\linewidth]{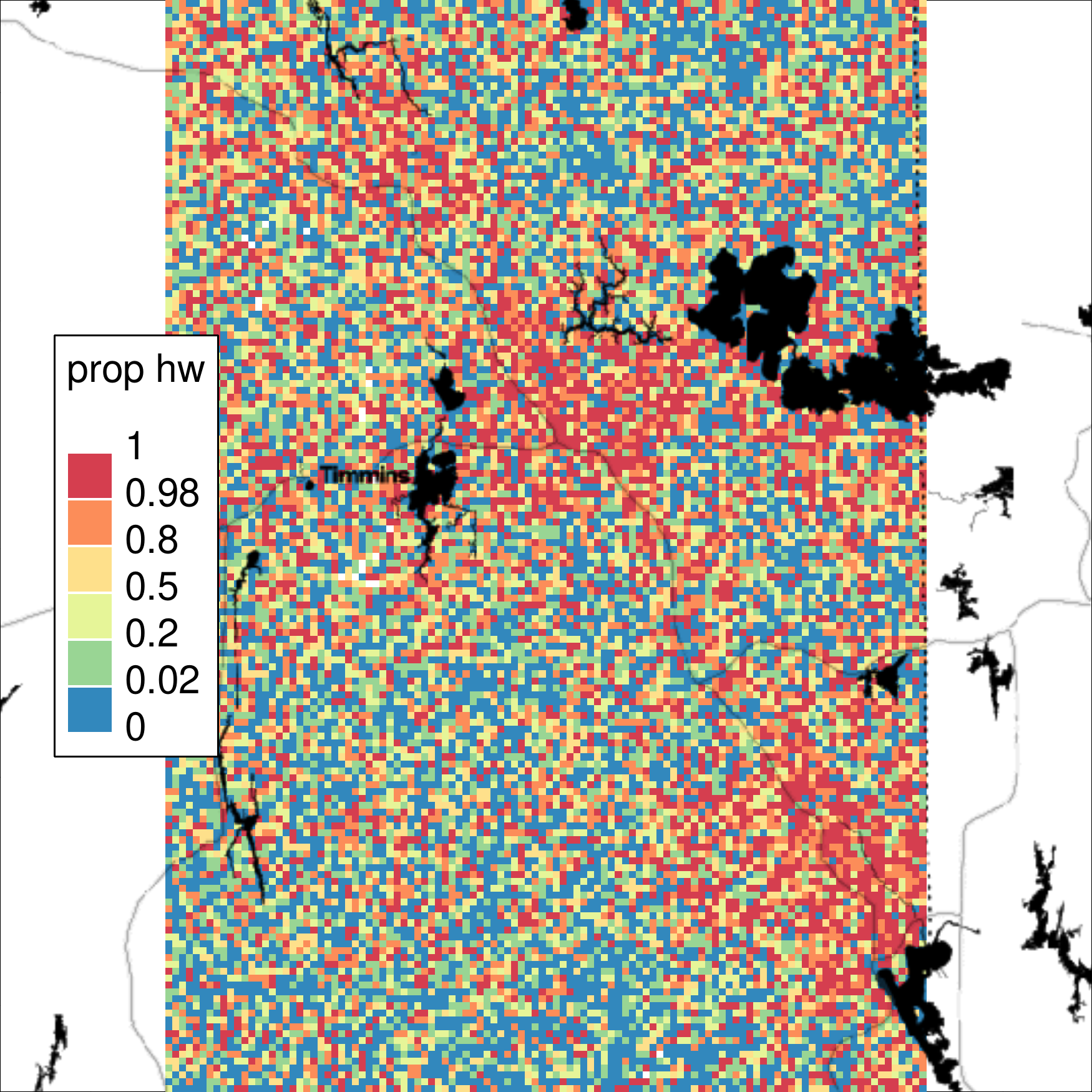}
  \caption{25 train - Sample 2}
  \label{25-2}
\end{subfigure}
\begin{subfigure}{.3\textwidth}
  \centering
  \includegraphics[width=0.88\linewidth]{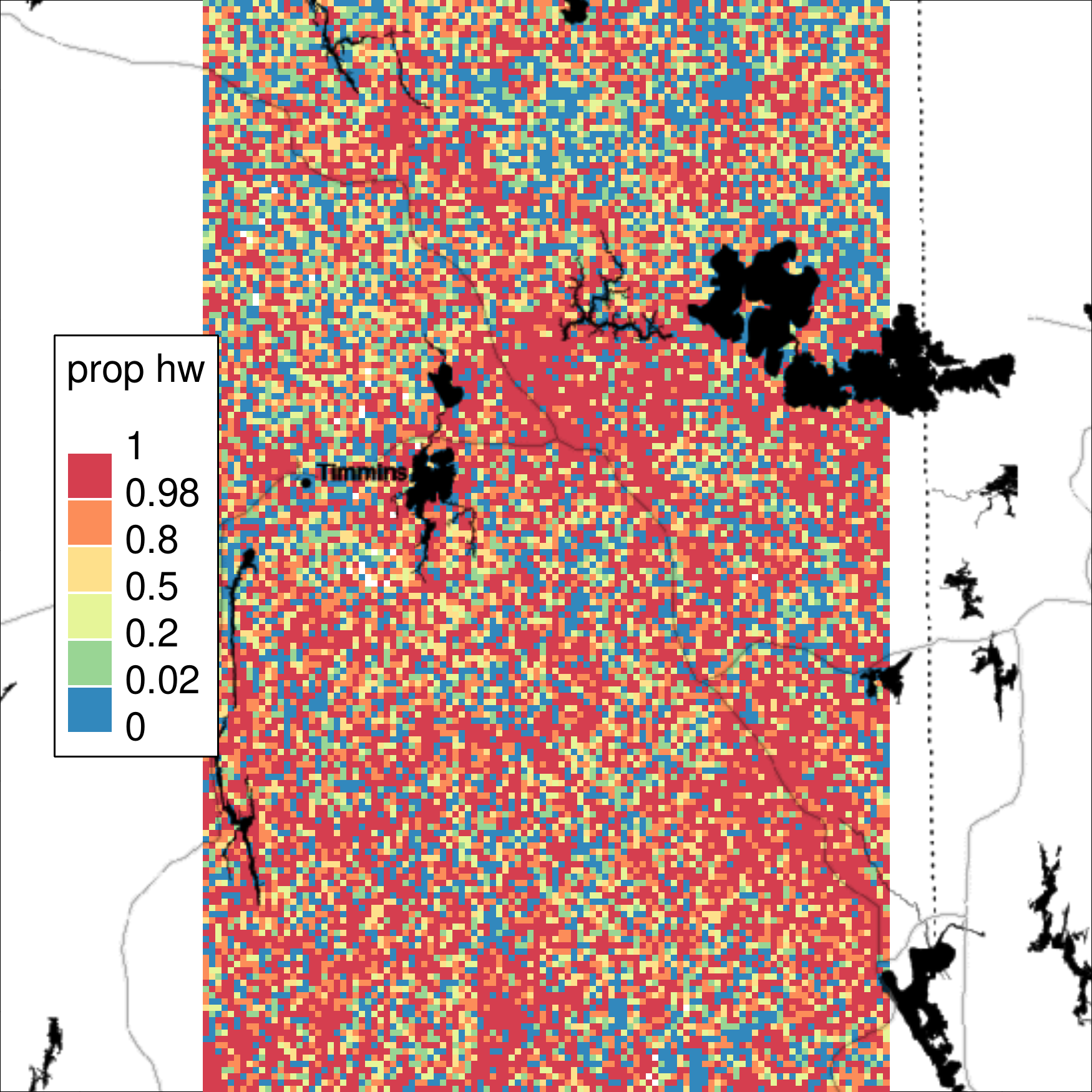}
  \caption{10 train - Sample 2}
  \label{25-3}
\end{subfigure}
\begin{subfigure}{.3\textwidth}
  \centering
  \includegraphics[width=0.88\linewidth]{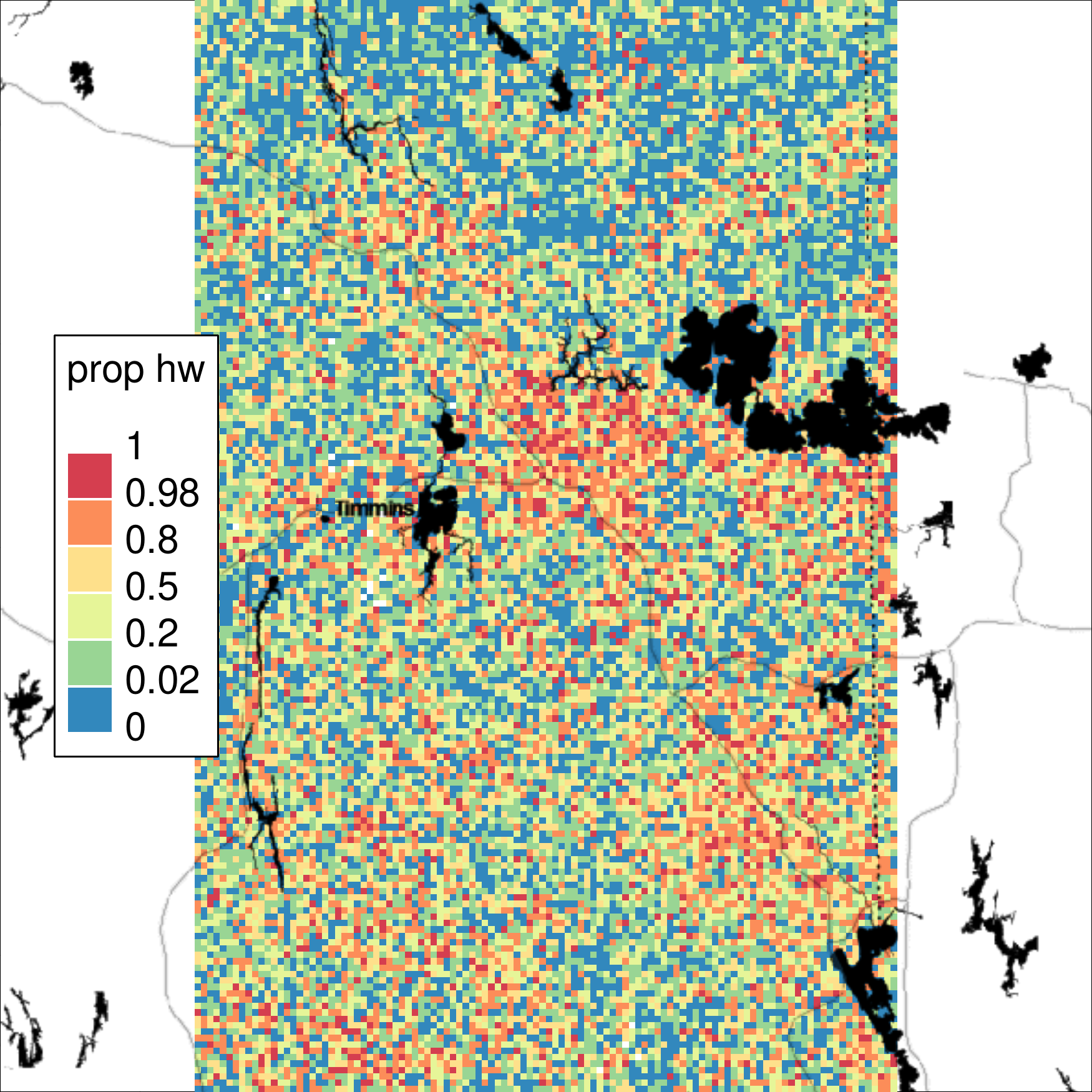}
  \caption{100 train - Sample 3}
  \label{15-1}
\end{subfigure}
\begin{subfigure}{.3\textwidth}
  \centering
  \includegraphics[width=0.88\linewidth]{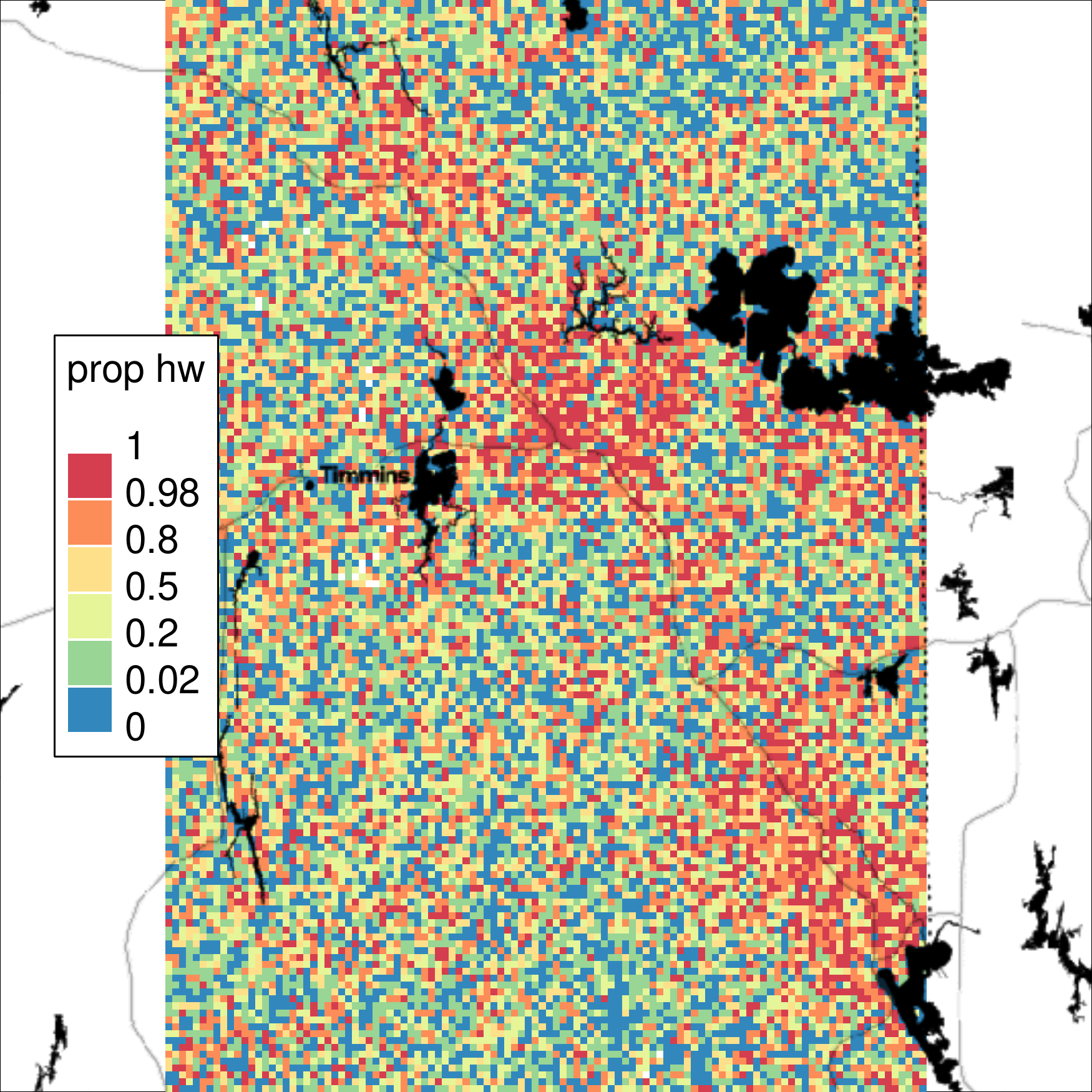}
  \caption{25 train - Sample 3}
  \label{15-2}
\end{subfigure}
\begin{subfigure}{.3\textwidth}
  \centering
  \includegraphics[width=0.88\linewidth]{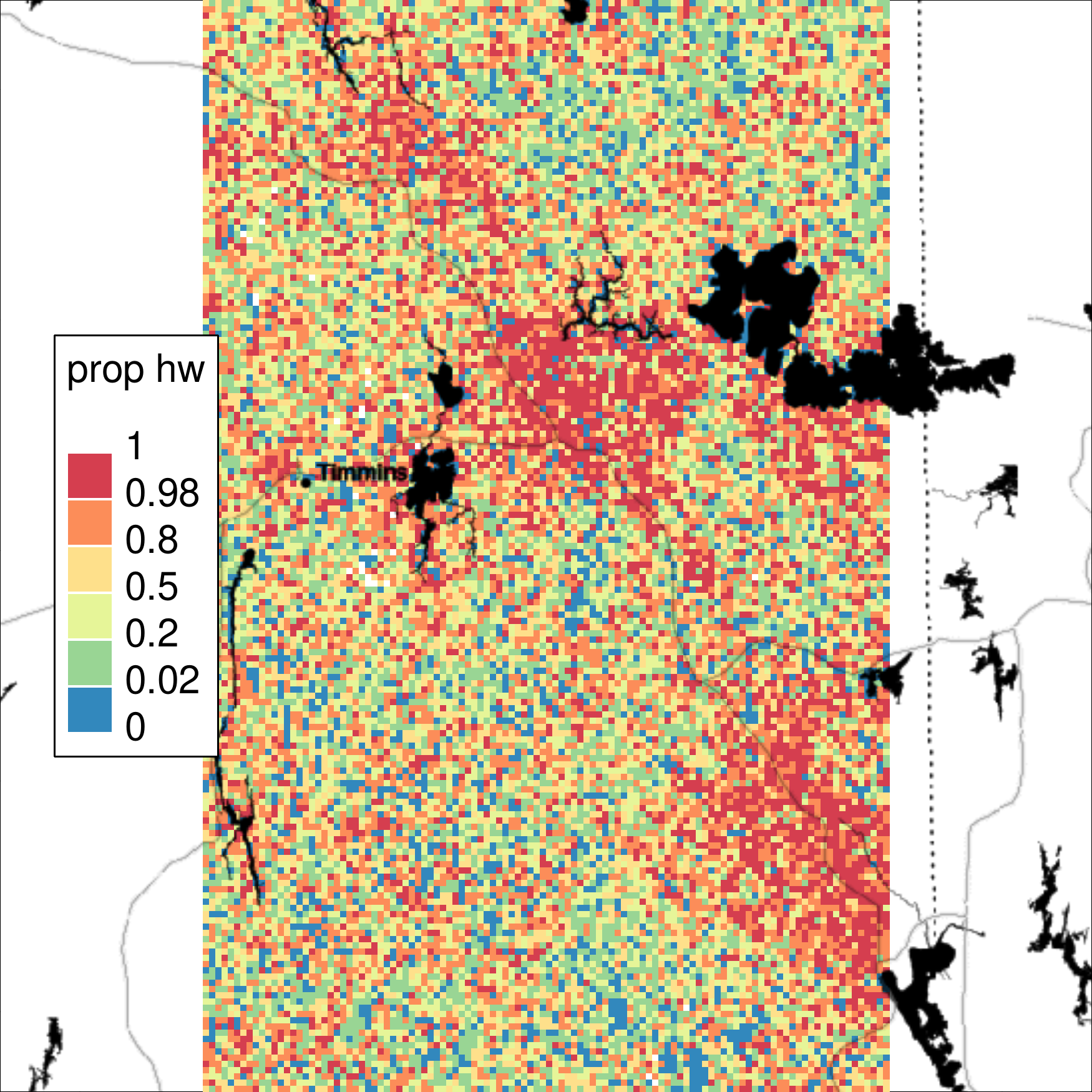}
  \caption{10 train - Sample 3}
  \label{15-3}
\end{subfigure}
\begin{subfigure}{.3\textwidth}
  \centering
  \includegraphics[width=0.88\linewidth]{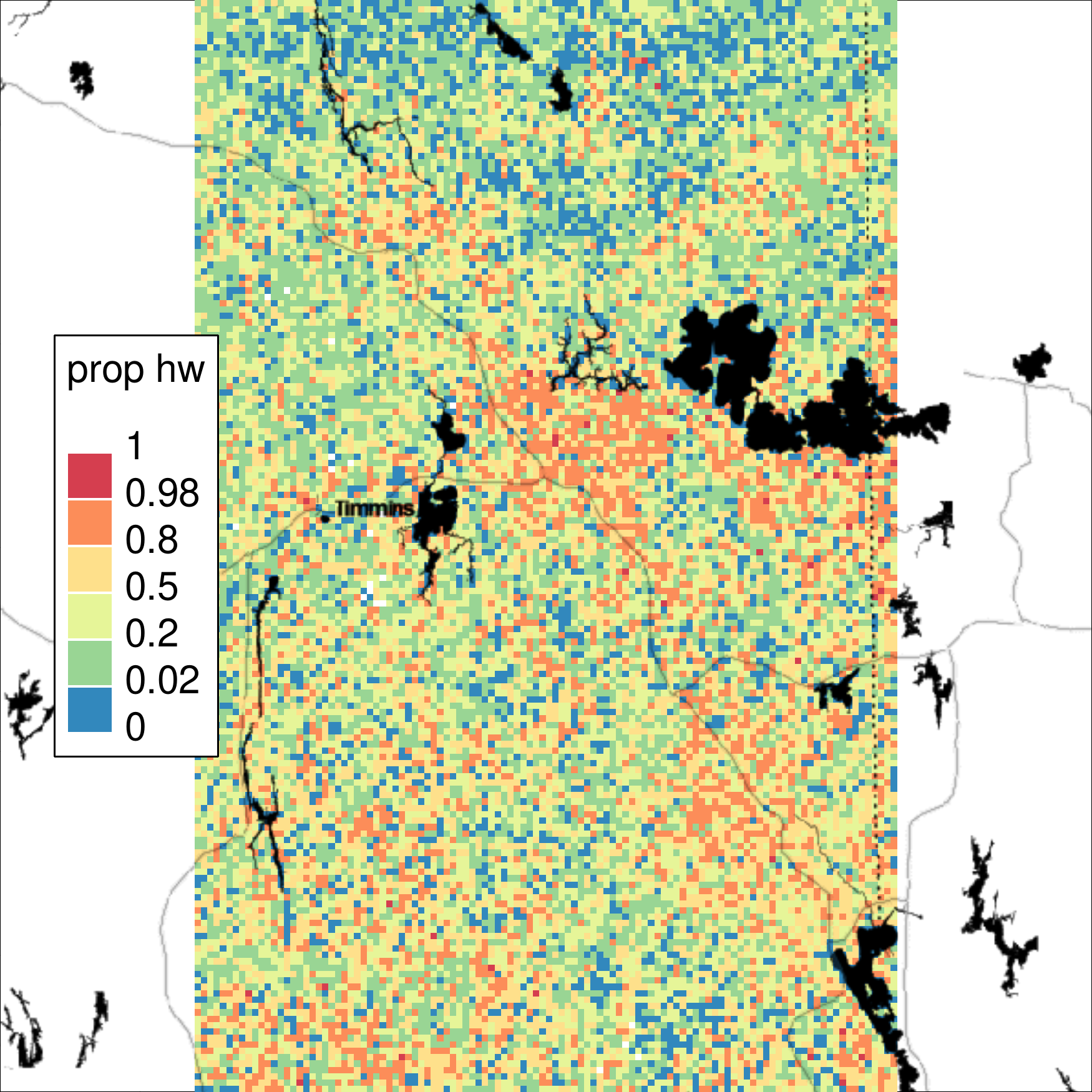}
  \caption{100 train - Sample mean}
  \label{100rho}
\end{subfigure}
\begin{subfigure}{.3\textwidth}
  \centering
  \includegraphics[width=0.88\linewidth]{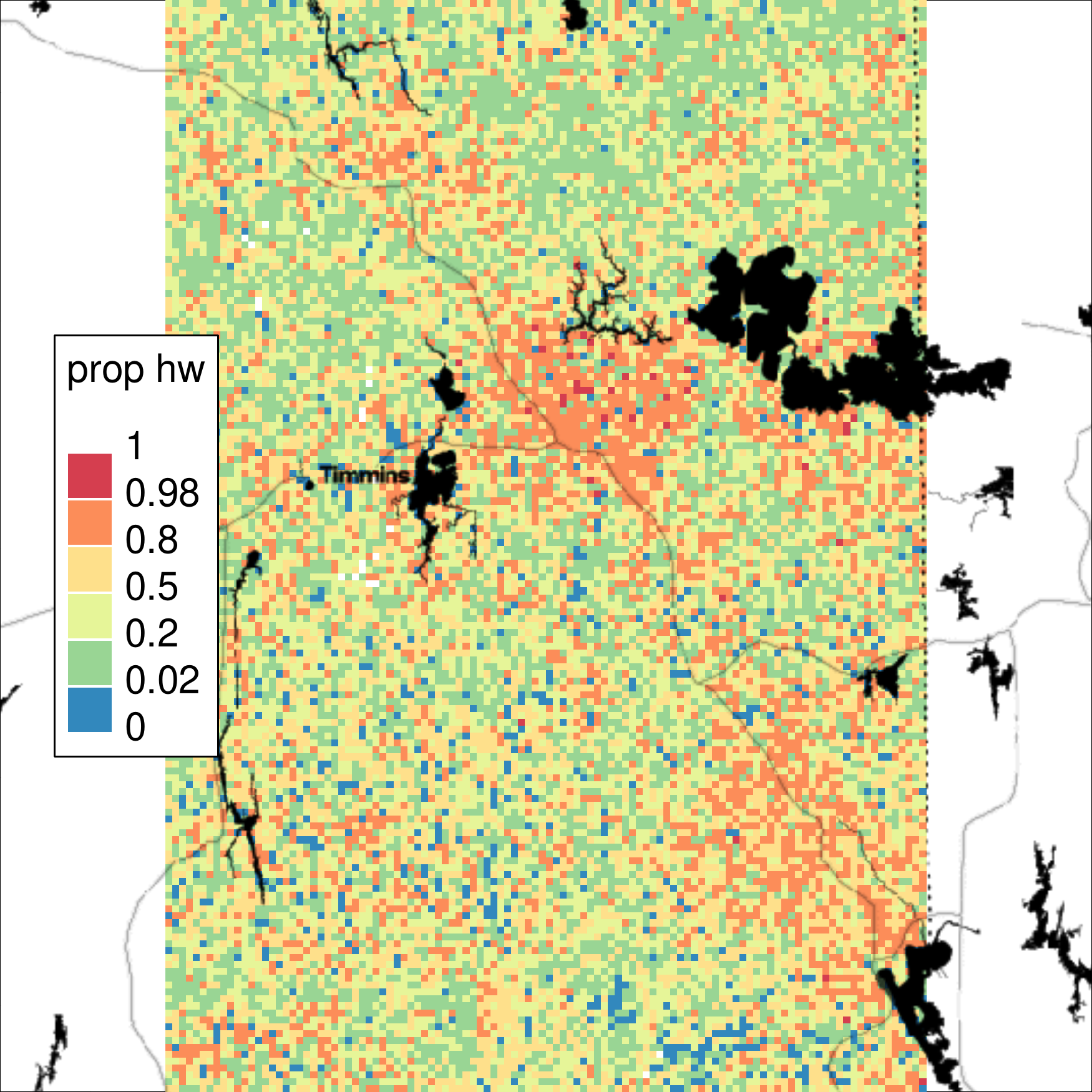}
  \caption{25 train - Sample mean}
  \label{100u}
\end{subfigure}
\begin{subfigure}{.3\textwidth}
  \centering
  \includegraphics[width=0.88\linewidth]{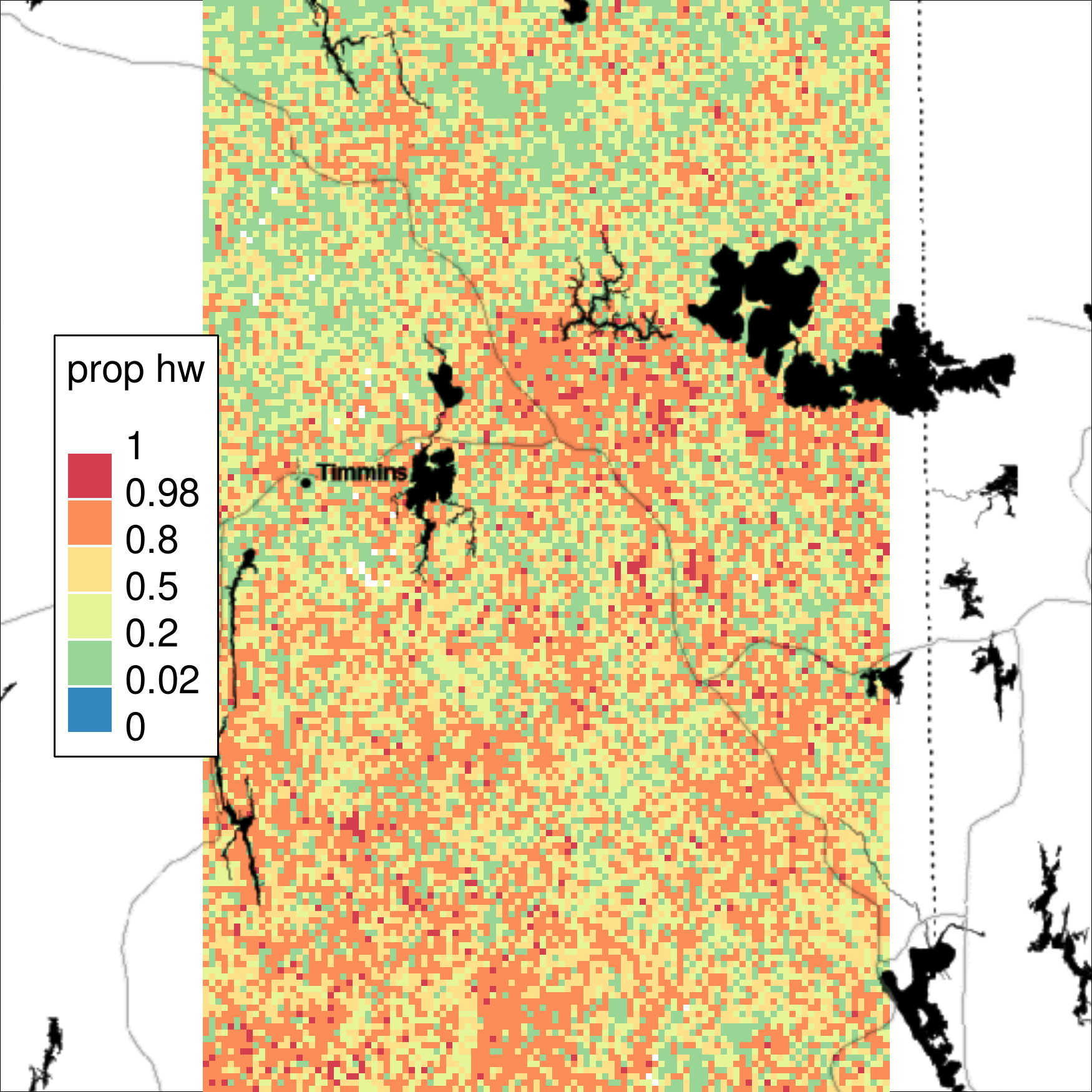}
  \caption{10 train - Sample mean}
  \label{25rho}
\end{subfigure}
\caption{Three posterior samples of the hardwood proportion surface $p(s^*)$ along with their posterior means from different training data sizes (Background \copyright \href{http://stamen.com}{Stamen Design}).}
\label{3RasterPost}
\end{figure}

Figure \ref{3RasterPost} shows images of three different posterior samples along with posterior means (in each row) generated from fitting different training data sizes. With fewer training data the posterior rasters appear to become smoother, possibly indicating less precise predictions. 



\begin{figure}[h]
\centering
\begin{subfigure}{.49\textwidth}
  \centering
  \includegraphics[width=1\linewidth]{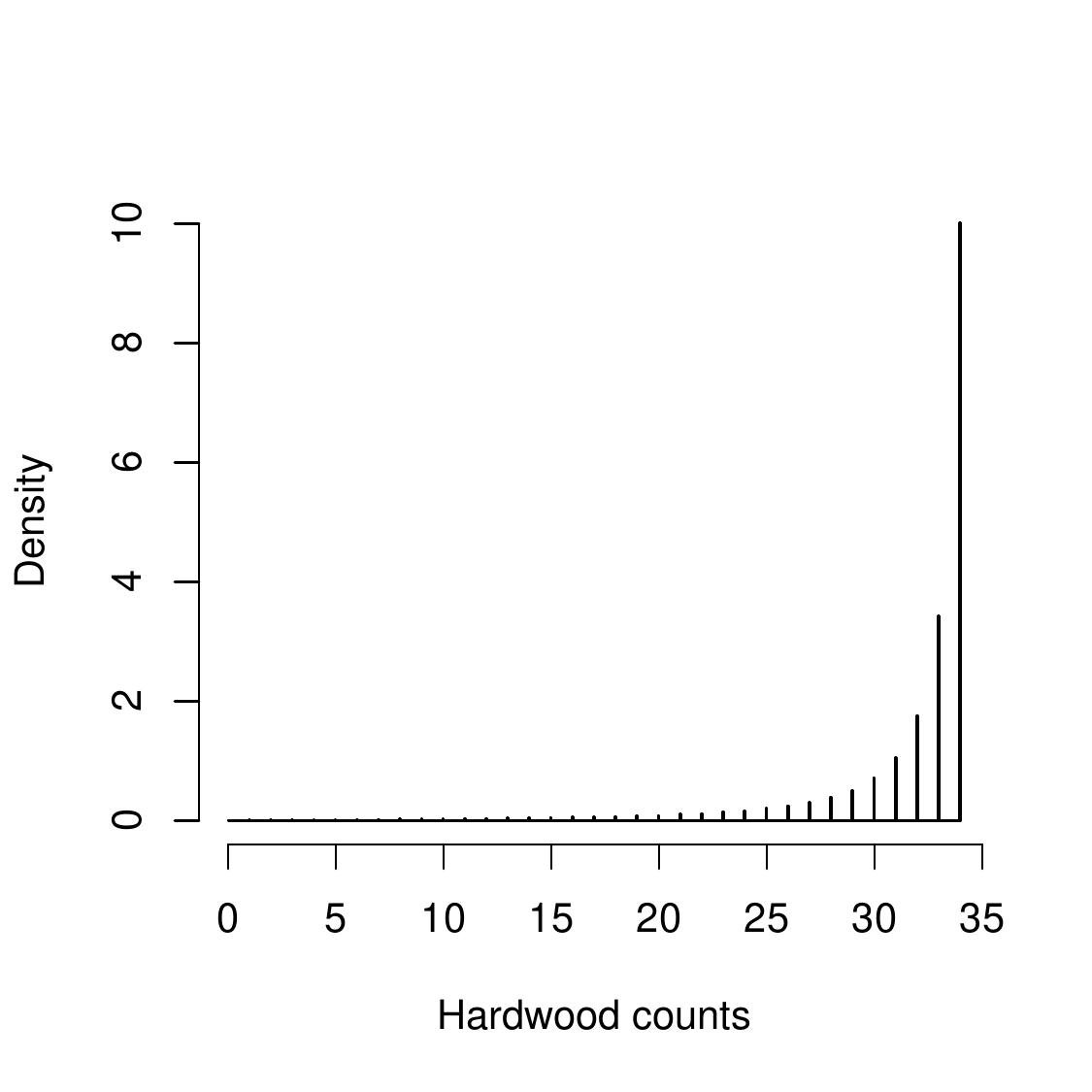}
  \caption{All hardwoods, plot 37}
  \label{100rho}
\end{subfigure}
\begin{subfigure}{.49\textwidth}
  \centering
  \includegraphics[width=1\linewidth]{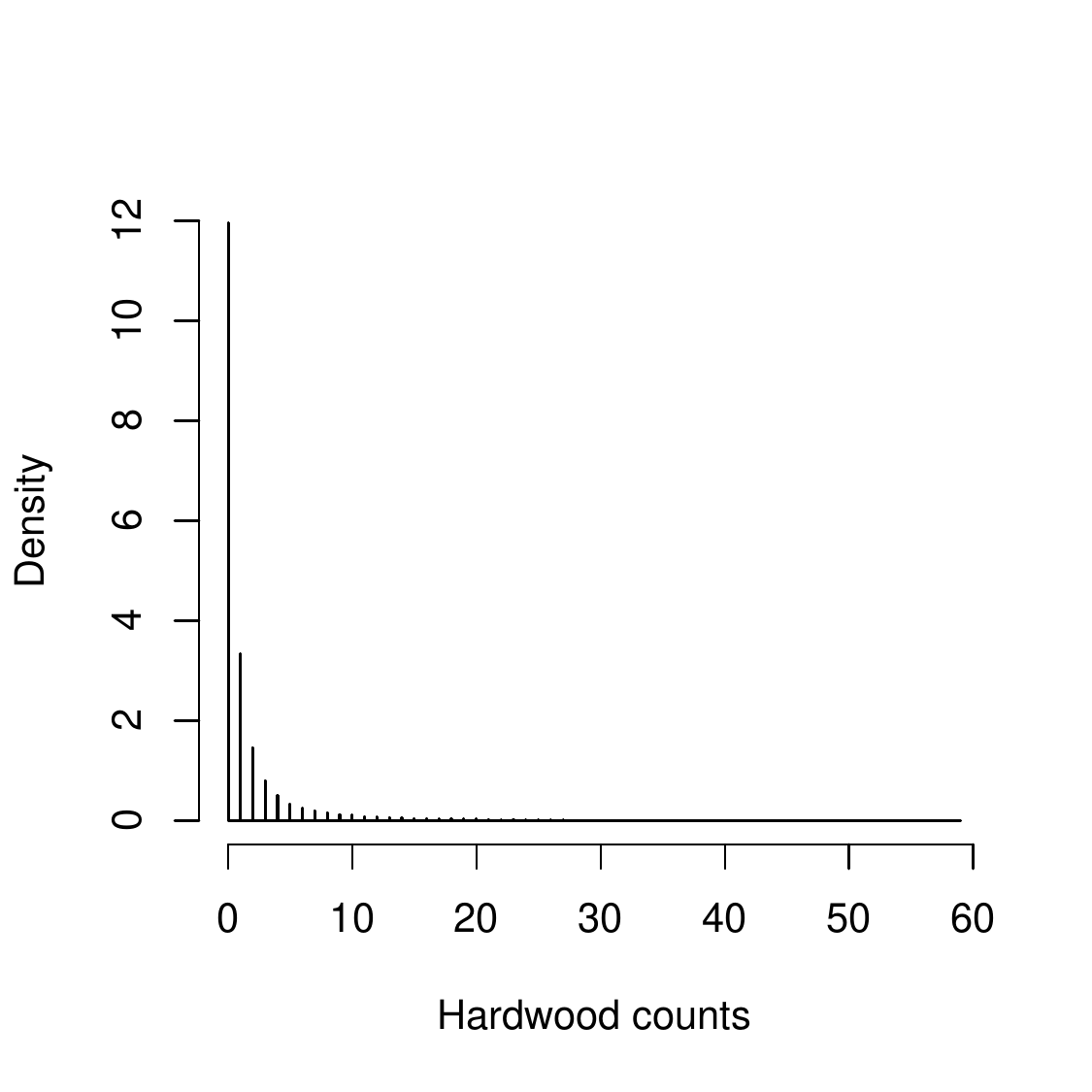}
  \caption{No hardwoods, plot 15}
  \label{100u}
\end{subfigure}
\caption{Posterior distributions of hardwood counts from two validation plots.}
\label{narrowCI_two}
\end{figure}

The 62 validation sites with their ground truth number of hardwood trees are used to evaluate predictions by summarizing results over all corresponding sites. The number of hardwood trees in each validation site is predicted and their coverage probabilities calculated from posterior intervals of hardwood counts. Table \ref{coverage} shows the corresponding coverage probabilities of 95\%, 80\%, and 50\% Posterior Credible Intervals (CI) for different training data size, averaged over five different simulations. Note that many observed proportions are 0 or 1, and the hardwood count posteriors will not be symmetric. To illustrate this, Figure \ref{narrowCI_two} shows the histograms of hardwood count posteriors for two validation plots where in one all are hardwoods and in the other none. We calculate the narrowest credible intervals for each validation plot, and compute their average coverages and widths as shown in Table \ref{coverage}.

\begin{table}[h]
\caption{Empirical Coverage of Posterior Credible Intervals and their Average Width. All results are averaged over 5 different simulations.}
\begin{tabular}{ |p{1.8cm}|p{1.8cm}|p{1.8cm}|p{1.8cm}|p{1.8cm}|p{1.8cm}|p{1.8cm}| }
 \hline
 & \multicolumn{3}{|c|}{Empirical Coverage of CI} & \multicolumn{3}{|c|}{Average CI Width}\\
 \hline
\#ofTrain & \textbf{95 \%} & \textbf{80 \%} & \textbf{50 \%} & \textbf{95 \%} & \textbf{80 \%} & \textbf{50 \%} \\ \hline \hline

100 Sites & 97 \% & 87 \% & 59 \% & 19.98 & 11.42 & 4.41 \\
25 Sites & 96 \% & 86 \% & 55 \% & 21.91  & 12.12 & 4.49 \\
10 Sites & 95 \% & 78 \% & 55 \% & 26.64 & 13.71 & 4.35 \\
 \hline
\end{tabular}
 \label{coverage}
\end{table}

The empirical coverage probabilities tend to exceed their theoretical values, meaning the intervals provided are on the conservative side. Overall, coverage probabilities are all at a desirable value. Table \ref{coverage} also includes the average width of the posterior intervals, which shows \emph{on average} wider intervals with fewer training data, as expected. 


\subsection{Comparison of BGLGM with Logistic Regression}\label{comp}

In this section we will show the difference in performances between the BGLGM and a simple Logistic Regression. Fitting a Logistic Regression model to this dataset using the function {\tt glm} in R is a frequentist way of analyzing this dataset, while BGLGM is a Bayesian approach. We will compare them both through their performance in point estimations via RMSEs, as well as their performance of distributions.

Table \ref{RMSE} is reporting the RMSEs of hardwood probabilities for the 62 validation sites, computed from runs with 100, 25, and 10 training data, for five different simulations. The RMSEs of BGLGM are calculated using posterior means. As it is observed, on average, RMSEs of BGLGM are smaller compared to Logistic Regression (GLM), indicating more accurate predictions. RMSEs increase with less ground truth data fitted to the model; verifying the results shown in the previous section.


\begin{table}
\centering
\caption{RMSE of predicted hardwood probabilities}
\begin{tabular}{ |p{2.5cm}|p{1.5cm}|p{1.5cm}|p{1.5cm}|p{1.5cm}|p{1.5cm}|p{1.5cm}|}
 \hline
 & \multicolumn{5}{|c}{RMSE} & \multicolumn{1}{|c|}{Avg RMSE}\\
 \hline
\#ofTrain & Sim 1 & Sim 2 & Sim 3 & Sim 4 & Sim 5 & \\ \hline \hline

100(BGLGM) &  0.2276 & 0.2072 & 0.2096 &  0.1983 & 0.1569 &  0.1999 \\
100(GLM) &  0.2333 & 0.2060 & 0.2137 & 0.1969 & 0.1575 & 0.2015 \\ 
&  &  &  &  &  & \\

25(BGLGM) &  0.2284 & 0.2106 & 0.2335 & 0.1985  & 0.2268 & 0.2196 \\
25(GLM) &  0.2493 & 0.2256 & 0.2455 & 0.2005  &  0.2569 & 0.2356 \\
&  &  &  &  &  & \\

10(BGLGM) &  0.2912 & 0.2435 & 0.3639 & 0.2080 & 0.2472 & 0.2708 \\
10(GLM) &  0.3509 & 0.2796 & 0.4805 & 0.2357 & 0.3168 & 0.3327 \\
 \hline
\end{tabular}
 \label{RMSE}
\end{table}

%
%
%

\begin{figure}[h]
\centering
\begin{subfigure}{.45\textwidth}
  \centering
  \includegraphics[width=1\linewidth]{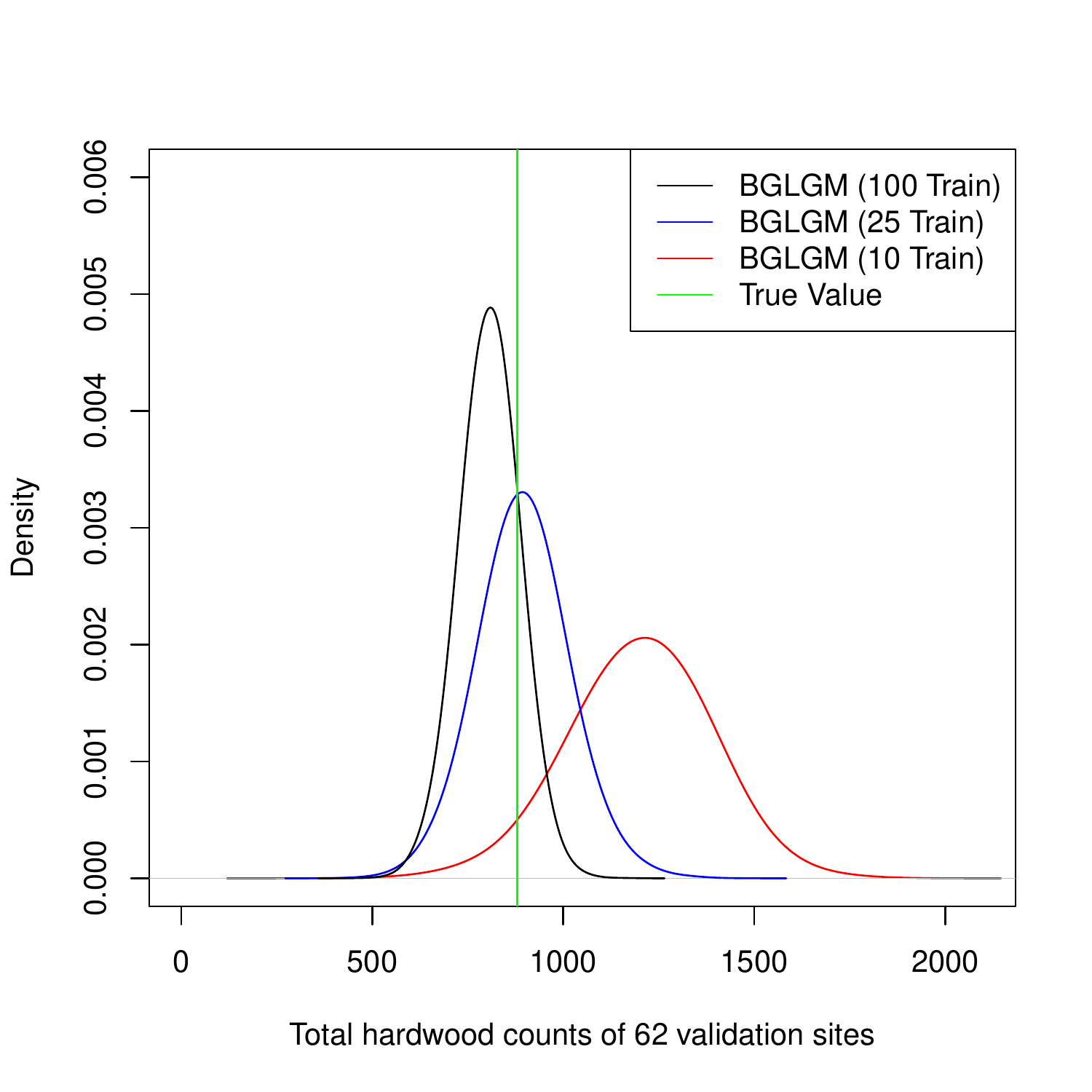}
  \caption{BGLGM}
  \label{dist_post}
\end{subfigure}
\begin{subfigure}{.45\textwidth}
  \centering
  \includegraphics[width=1\linewidth]{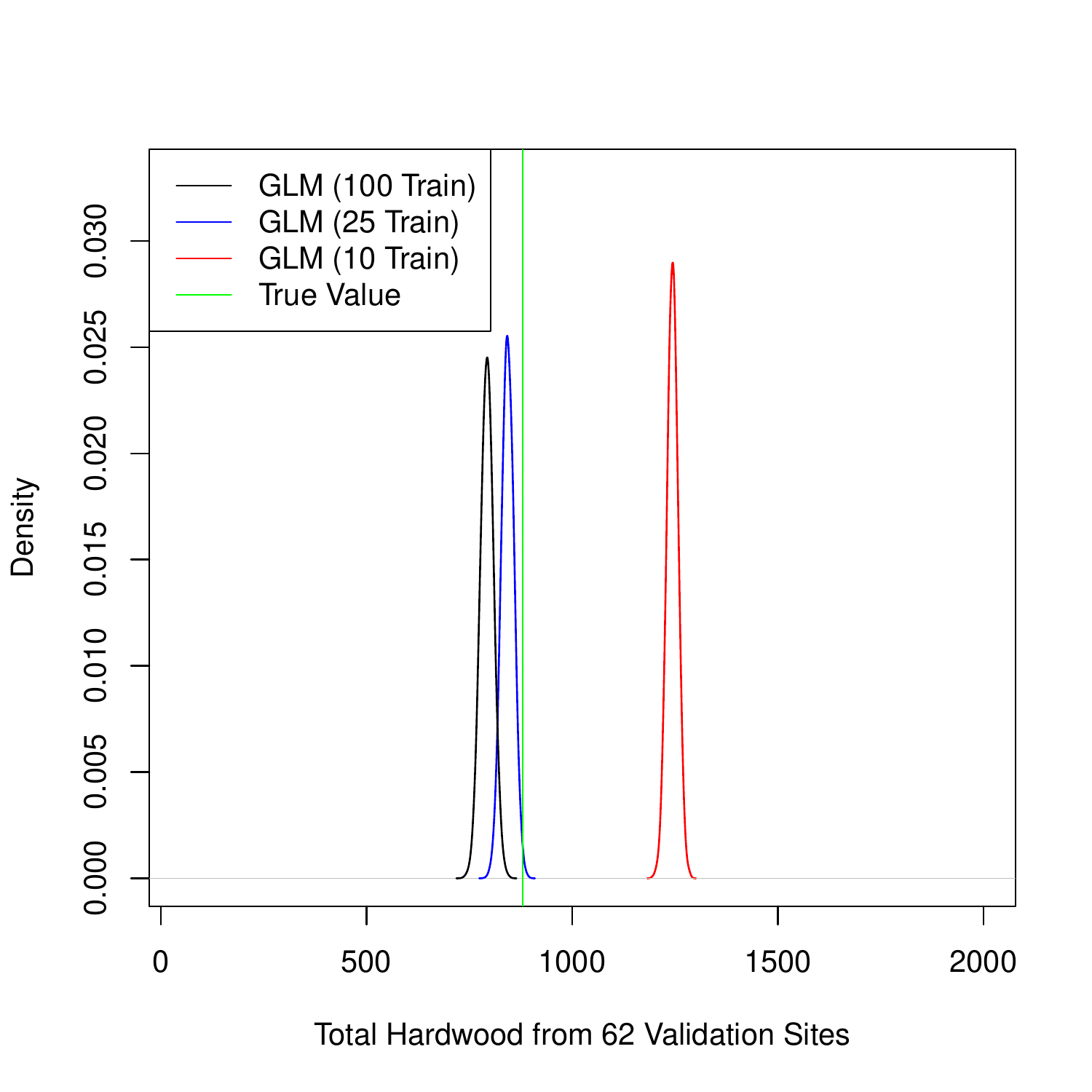}
  \caption{GLM}
  \label{dist_glm}
\end{subfigure}
\caption{Comparing BGLGM posterior distributions of total number of hardwood trees to the frequentist distributions from GLM.}
\label{sim_post_glm}
\end{figure}

To compare the predictive distributions of the GLM and BGLGM, we simulated 10,000 hardwood counts for each validation site using the estimated probabilities from GLM. These are compared to 10,000 MCMC posterior samples from the BGLGM using the distributions of the \emph{total} hardwood counts from all 62 validation sites. Figure \ref{sim_post_glm} is showing the corresponding distributions from BGLGM and GLM for only the first simulation. The distributions from GLM are significantly narrower compared to the ones from BGLGM, as should be expected, since the GLM is ignoring errors in the parameter estimates. The BGLGM posterior distributions with all training data sizes capture the true value shown in green within their 95\% intervals, while GLM with 10 and even 100 training data points fails to do so. In addition, from Figure \ref{dist_post}, we also observe that the posterior distributions become wider with less training data as expected. In conclusion, the BGLGM is a more reliable method compared to the GLM, in terms of both prediction accuracy and the ability of explaining uncertainties. Note that this process has been repeated for four other simulations with figures shown in the Appendix. 

Overall, a Bayesian approach is more reliable compared to a frequentist approach, since more types of uncertainty are taken into account. The simple Logistic Regression has artificially narrow prediction intervals, while BGLGM includes the true value within its 95\% intervals for this dataset. 

\subsubsection{Stratified Sampling of training data}\label{stratified}


%

\begin{figure}[h]
\centering
\begin{subfigure}{.45\textwidth}
  \centering
  \includegraphics[width=1\linewidth]{Figures/sim1_post_totHard.pdf}
  \caption{Random}
  \label{random}
\end{subfigure}
\begin{subfigure}{.45\textwidth}
  \centering
  \includegraphics[width=1\linewidth]{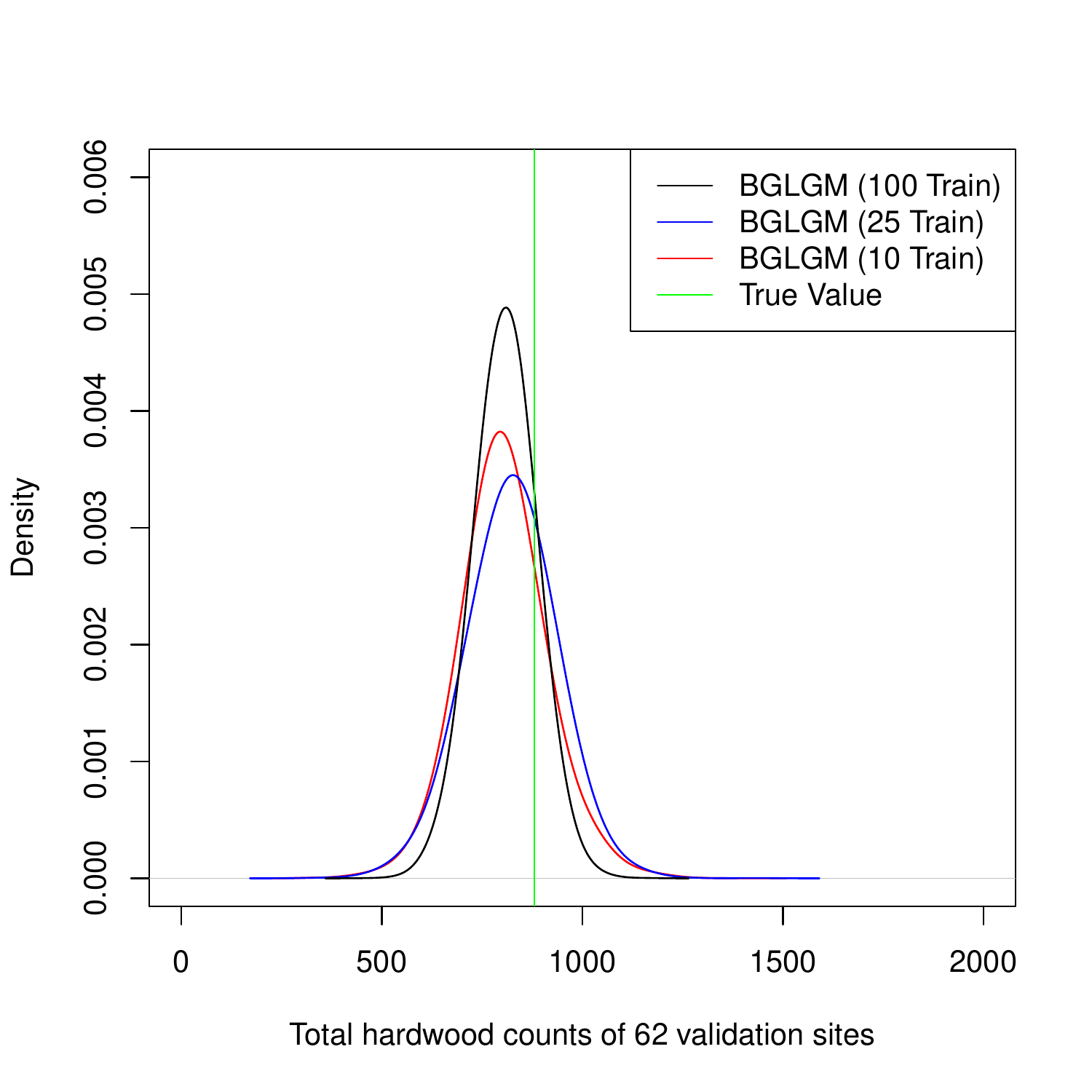}
  \caption{Stratified}
  \label{stratpost}
\end{subfigure}
\caption{Comparing random vs stratified sampling for total hardwood posterior distributions.}
\label{strat-random}
\end{figure}

\begin{figure}[h]
\centering
  \includegraphics[height = 10cm, width=11cm]{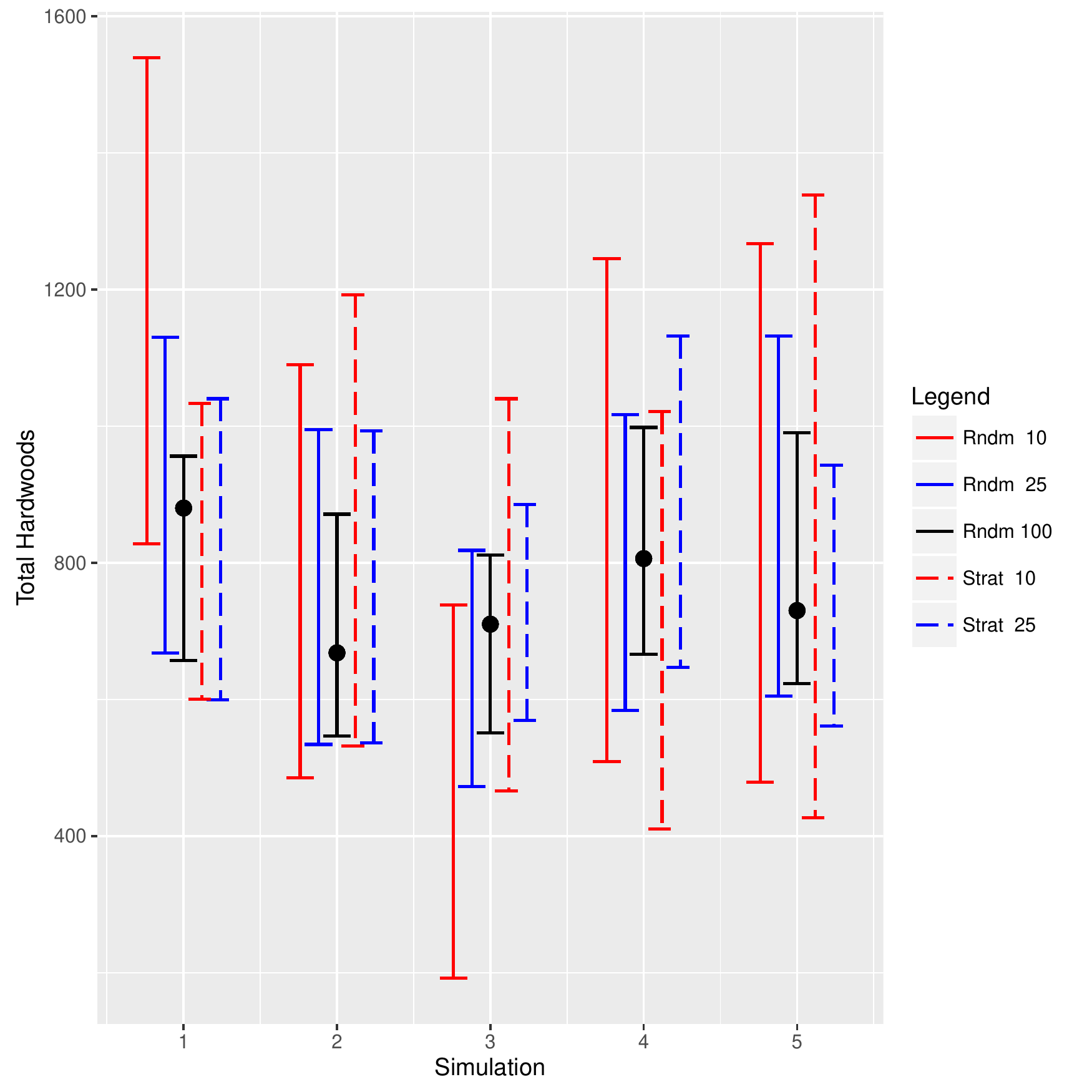}
\caption{95\% posterior intervals of Random sampling vs Stratified sampling from all five simulations.}
\label{allsim_strat}
\end{figure}

Figure \ref{strat-random} compares the posterior distributions of the total hardwood trees from both random sampling and stratified sampling on the first of five simulations. Prediction intervals from all five simulations are shown in Figure \ref{allsim_strat}.
The posterior distributions all contain the true value within their 95\% posterior interval, however the uncertainty is generally less under stratified sampling in most cases. In Figure \ref{strat-random} it is notable that the stratified posterior with 10 training data contains the true value near its mode, while with the random posterior it is covered around the tail area.

In simulations 2 and 4 results are roughly comparable, while in simulation 3 the stratified posterior with 10 data points captures the true value around its mode. On the other hand, in simulation 5, the stratified posterior with 10 data points becomes more dispersed while with 25 more narrow. Overall, the stratified sampling approach shows only \emph{potential} in improving results and thus may not be of significant improvements.

More details on the MCMC trace plots and the posterior distributions of each model parameter for the stratified sampling is included in the Appendix (for all simulations). 

\section{Discussion}\label{discussion}

In this paper, we have analyzed the spatial data from the Timiskaming and Abitibi River forests in Ontario, Canada. We have studied the prediction of hardwood tree counts from elevation and vegetation index. We implemented a bespoke MCMC algorithm for posterior simulation of a Bayesian Generalized Linear Geostatistical Model (BGLGM), in order to make spatial predictions for new sites in the forests. The bespoke MCMC performed well with this dataset while the general purpose ``PrevMap" package struggled. We compared the Bayesian model with the frequentist Logistic Regression model. Although the dataset is imbalanced and contains many zero hardwood counts, the Bayesian approach provided unbiased estimates with reasonable uncertainties, while the overly simplistic Logistic Regression underestimated the uncertainty associated with the predictions. More importantly, with ground truth data as small as 10 points, BGLGM captured the true value of hardwood tree counts within its 95\% posterior intervals, while the Logistic Regression failed even with 100 training points. This suggests with fewer ground truth data collected and hence reduction in expenses, good estimates of hardwood counts are still present and can capture the true value but perhaps with more uncertainties involved. This result is fairly important in terms of saving time and money for companies to gather such data.

Furthermore, a stratified sampling strategy of choosing the subset training data showed potential improvements in terms of predictions and uncertainties. However, these improvements are not always guaranteed.  


As future work, one can further extend this model for multiple forests, where forests with similar features are considered to have high correlation indicated within priors and hence facilitate future spatial predictions for similar forests. This will significantly help reduce the redundant collection of data from similar forests.

\section*{Acknowledgements}

The authors thank Philip E. J. Green, M.Sc., and the First Resource Management Group for introducing us to this problem, providing scientific and technical advice, and making the data available.

Figures \ref{fig:forests_pic} and \ref{elev-strat} have cartography by \href{http://www.nrcan.gc.ca/earth-sciences/geography/topographic-information/free-data-geogratis/geogratis-web-services/17216}{The Canada Base Map --- Transportation (CBMT) web mapping services of the Earth Sciences Sector (ESS) at Natural Resources Canada (NRCan)} licensed as the \href{http://open.canada.ca/en/open-government-licence-canada}{Open Government Licence --- Canada}.

Figures \ref{fig:timisk} and \ref{3RasterPost} have map tiles by \href{http://stamen.com}{Stamen Design} under \href{http://creativecommons.org/licenses/by/3.0}{CC BY 3.0}. Data by \href{http://openstreetmap.org}{OpenStreetMap}  available under the \href{http://opendatacommons.org/licenses/odbl}{Open Database License}.

The second and third authors hold Discovery grants from the Natural Sciences and Engineering Research Council of Canada.

\newpage

\bibliography{Forestry_paper}
\newpage
\appendix
\counterwithin{figure}{section}

\end{document}